\documentclass[fleqn,usenatbib]{mnras}

\usepackage{newtxtext,newtxmath}

\usepackage[T1]{fontenc}

\usepackage{ae,aecompl}

\usepackage{graphicx}
\usepackage{amssymb}
\usepackage{amsmath}

\newcommand{\mpro}{m_{\rm p}}

\newcommand{\cm}{\,{\rm cm}}

\newcommand{\kpc}{\,{\rm kpc}}
\newcommand{\Mpc}{\,{\rm Mpc}}

\newcommand{\s}{\,{\rm s}}

\newcommand{\yr}{\,{\rm yr}}

\newcommand{\Gyr}{\,{\rm Gyr}}
\newcommand{\erg}{\,\rm erg}

\newcommand{\K}{\,{\rm K}}

\newcommand{\msun}{\,{\rm M_{\odot}}}
\newcommand{\zsun}{\,{\rm Z_{\odot}}}
\newcommand{\zism}{{Z_{\rm MZR}}}

\newcommand{\tadv}{t_{\rm flow}}
\newcommand{\cs}{c_{\rm s}}
\newcommand{\vc}{v_{\rm c}}

\newcommand{\Mdotcrit}{\Mdot_{\rm crit}}

\newcommand{\mach}{\mathcal{M}}
\newcommand{\Bernoulli}{\mathcal{B}}

\newcommand{\Mstar}{M_*}
\newcommand{\Mdot}{{\dot M}}
\newcommand{\nH}{n_{\rm H}}

\newcommand{\kms}{\,\rm km\ s^{-1}}

\newcommand{\vr}{v_{r}}
\newcommand{\vphi}{v_{\phi}}

\newcommand{\Rvir}{R_{\rm vir}}
\newcommand{\Tvir}{T_{\rm vir}}

\newcommand{\Rsonic}{R_{\rm sonic}}

\newcommand{\der}{{\rm d}}

\newcommand{\tcool}{t_{\rm cool}}
\newcommand{\Rcool}{R_{\rm cool}}

\newcommand{\Rhalf}{R_{1/2}}

\newcommand{\thubble}{t_{\rm H}}

\newcommand{\tff}{t_{\rm ff}}

\newcommand{\tflow}{\tadv}

\newcommand{\Mhalo}{M_{\rm halo}}

\newcommand{\fb}{f_{\rm b}}

\newcommand{\Deltac}{\Delta_{\rm c}}
\newcommand{\Hnorm}{E(z)}
\newcommand{\Hnormp}[1]{E^{#1}(z)}
\newcommand{\Omo}{\Omega_{\rm m,0}}

\newcommand{\Rcirc}{R_{\rm circ}}
\newcommand{\rvir}{R_{\rm vir}}
\newcommand{\vvir}{v_{\rm vir}}

\newcommand{\SFR}{{\rm SFR}}

\newcommand{\Mgas}{M_{\rm gas}}
\newcommand{\Mthres}{M_{\rm thres}}

\newcommand{\sigrho}{\langle\delta\rho/\rho\rangle_{\rm rms}}

\newcommand{\avgmach}{\langle \log \mach\rangle}
\newcommand{\fvc}{f_{\vc}}

\title[The maximum accretion rate of hot gas]{The maximum accretion rate of hot gas in dark matter halos}

\author[J. Stern et al.]{
Jonathan Stern,$^{1}$\thanks{CIERA Fellow}\thanks{E-mail: jonathan.stern@northwestern.edu}
Drummond Fielding,$^{2}$
Claude-Andr{\'e} Faucher-Gigu{\`e}re$^{1}$
\newauthor 
and Eliot Quataert$^{3}$
\\
$^{1}$Department of Physics and Astronomy and CIERA, Northwestern University, Evanston, IL, USA\\
$^{2}$Center for Computational Astrophysics, Flatiron Institute, 162 5th Ave, New York, NY 10010, USA\\
$^{3}$Astronomy Department and Theoretical Astrophysics Center, University of California Berkeley, Berkeley, CA 94720, USA
}

\date{Accepted XXX. Received YYY; in original form ZZZ}

\pubyear{2019}

\begin{document}
\label{firstpage}
\pagerange{\pageref{firstpage}--\pageref{lastpage}}
\maketitle

\begin{abstract}
We revisit the question of `hot mode' versus `cold mode' accretion onto galaxies using steady-state cooling flow solutions and idealized 3D hydrodynamic simulations.
We demonstrate that for the hot accretion mode to exist, the cooling time is required to be longer than the free-fall time near the radius where the gas is rotationally-supported, $\Rcirc$, i.e.~the existence of the hot mode depends on physical conditions at the galaxy scale rather than on physical conditions at the halo scale.  
When allowing for the depletion of the halo baryon fraction relative to the cosmic mean, the longer cooling times imply that a virialized gaseous halo may form in halo masses below the threshold of $\sim10^{12}\msun$ derived for baryon-complete halos. We show that for any halo mass there is a maximum accretion rate for which the gas is virialized throughout the halo and can accrete via the hot mode of $\Mdotcrit\approx 0.7(\vc/100\kms)^{5.4}(\Rcirc/10\kpc)(Z/\zsun)^{-0.9}\msun\yr^{-1}$, where 
$Z$ and $\vc$ are the metallicity and circular velocity measured at $\Rcirc$. For accretion rates $\gtrsim\Mdotcrit$ the volume-filling gas phase can in principle be `transonic' -- virialized in the outer halo but cool and free-falling near the galaxy. 
We compare $\Mdotcrit$ to the average star formation rate (SFR) in halos at $0<z<10$ implied by the stellar-mass halo-mass relation. For a plausible metallicity evolution with redshift, we find that $\SFR \lesssim\Mdotcrit$ at most masses and redshifts, suggesting that the SFR of galaxies could be primarily sustained by the hot mode in halo masses well below the classic threshold of $\sim10^{12}\msun$.
\end{abstract} 


\section{Introduction}

The dynamics of the volume-filling gas phase in dark matter halos, and the nature of its accretion onto the galaxy, crucially depend on whether the cooling time $\tcool$ of virialized gas is longer or shorter than the free-fall time $\tff$. Since $\tff$ roughly equals the sound-crossing time in virialized gas, if $\tcool>\tff$ then the volume-filling phase can be quasi-static, supported against gravity by thermal pressure. Galaxy accretion in this regime is gradual and regulated by energy losses to radiation. 
In contrast if $\tcool<\tff$ then the rapid cooling prevents the formation of a pressure-supported gaseous halo, and the halo gas free-falls onto the galaxy. 
These two distinct regimes for the nature of galaxy accretion, known respectively as `hot mode' and 'cold mode' accretion, were originally discussed by \cite{WhiteRees78} who demonstrated that the ratio $\tcool/\tff$ increases with halo mass $\Mhalo$. They identified a threshold mass scale of $\Mthres\sim 10^{12}\msun$ where $\tcool\sim\tff$, similar to the threshold previously derived for self-gravitating gas clouds (\citealt{ReesOstriker77, Silk77}). 
\citeauthor{BirnboimDekel03} (2003, hereafter BD03) and \cite{DekelBirnboim06} later connected these two regimes to the stability of the virial shock. 
Using analytic arguments and 1D simulations, they demonstrated that the rapid cooling of postshock gas in low mass halos leads to an unstable shock, so gas accreting from the intergalactic medium (IGM) remains cool ($\sim10^4\K$) and free-falling down to the galaxy scale. 
Once however $\Mhalo$ surpasses $\Mthres\sim10^{11.5}\msun$ the conditions for a stable shock are met at the galaxy scale, and a shock forms and expands into the halo heating the volume-filling phase to the virial temperature $\Tvir$. 

More recently \cite{Fielding+17} used idealized 3D simulations to study how the two regimes for the halo gas are affected by kinetic feedback from stars. They demonstrated that in the $\Mhalo>\Mthres$ regime the outflows are confined by the hot halo gas, and the physics of the volume-filling phase are similar to that suggested by BD03. 
In low mass halos however the effect of feedback is much more dramatic -- galaxy outflows shock against IGM inflows at halo radii, well beyond the radius where the shock initially forms in the BD03 simulations. In this regime the halo gas forms a multi-phase medium dominated by turbulence and bulk inflows/outflows. 

A considerable effort has been devoted to detecting these two regimes for galaxy accretion, and the transition between them, in cosmological simulations 
(e.g., \citealt{Keres+05,Keres+09,Keres+12,Birnboim+07,Ocvirk+08,Brooks+09,Oppenheimer+10,FaucherGiguere+11,vandeVoort+11,Nelson+13,Correa+18}). To discriminate between gas which has shocked prior to accretion and gas which has not shocked, most of these studies identified the maximum temperature $T_{\rm max}$ a fluid element had reached before accreting onto the galaxy. In studies where the `hot' and `cold' accretion modes are differentiated by a constant cut in temperature $T_{\rm cut}\approx10^{5.5}\K$ the gas was found to be entirely cold below $\Mhalo\approx10^{11.5}\msun$, consistent with the conclusion of BD03  (e.g.~\citealt{Keres+05}). This trend however could be driven by $\Tvir$ dropping below $T_{\rm cut}$ in low mass halos, in which case even virial-temperature gas would be classified as `cold', as discussed in \cite{Nelson+13} and acknowledged by many of the cited studies. \cite{Nelson+13} scaled $T_{\rm cut}$ with  $\Tvir$ and found that the hot accretion mode is present even in halo masses well below $10^{11.5}\msun$ (see also figure~8 in \citealt{vandeVoort+11}), in contrast with the 1D simulations of BD03. However, given that even in the free-fall regime inflows potentially shock and reach a temperature $\sim\Tvir$ due to the interaction with outflows as seen in the simulations of \cite{Fielding+17}, $T_{\rm max}$ may not be a good discriminator between the two regimes. An alternative method to distinguish between gradual, pressure-supported accretion and supersonic free-fall in cosmological simulations would thus be useful.

Another complication arises since $\tcool$ depends on the gas density, which implies that the transition between cold and hot mode accretion depends on the gas mass $\Mgas$ available to form the hot volume-filling phase. The idealized studies mentioned above assumed $\Mgas$ roughly equals the cosmic halo baryon budget $\fb\Mhalo$ ($\fb\approx0.16$ is the cosmic baryon fraction). If however a significant fraction of halo baryons are confined to filaments and subhalos, or, alternatively, if the halo baryons were ejected from the halo by unbound galaxy outflows at earlier times, then $\Mgas$ will be lower than $\fb\Mhalo$, $\tcool$ would correspondingly be longer, and the transition to pressure support would occur in halos less massive than derived by assuming halos are baryon-complete. Specifically, there is mounting observational evidence for the existence of strong unbound outflows, especially in dwarf galaxies which reside in halos with mass lower than $\Mthres$ (e.g.~\citealt{HeckmanThompson17, Chisholm+17}). 
Also, cosmological simulations which model galaxy outflows often predict halo gas masses lower than $\fb\Mhalo$. 
In the FIRE zoom-in simulations \cite{Hafen+19} find a baryon mass of $\approx0.3\fb\Mhalo$ in $\sim10^{11}\msun$ halos at low redshift.
In lower mass $\sim 10^{10}\msun$ halos in FIRE the baryon fraction is even lower, less than $10\%$ of the cosmic baryon budget. 
A low baryon mass fraction in low mass halos is found also in the EAGLE cosmological simulations (\citealt{Davies+19,Oppenheimer+19}).\footnote{In contrast, the baryon fraction in dwarf halos in the IllustrisTNG simulations appears to be closer to the baryon budget (\citealt{Nelson+18}).} 
Thus, both observations and some theoretical studies suggest that $\Mgas$ could be well below $\fb\Mhalo$ in halos with mass below the threshold derived assuming $\Mgas\approx\fb\Mhalo$, in which case hot mode accretion could be important also in low mass halos.

In this work (Paper II) we deduce the conditions under which hot mode accretion is possible by analyzing the properties of cooling flow solutions. Cooling flows were originally discussed in the context of gas in the centers of clusters (\citealt{MathewsBregman78,Cowie+80,Fabian+84,Bertschinger89}), and adapted to galaxy scale halos in the first paper in this series (\citealt{Stern+19}, hereafter Paper I). Here, we focus on halo masses which are comparable or below the classic threshold for the formation of a hot halo $\Mthres\sim10^{12}\msun$. We demonstrate that in low mass halos hot mode accretion depends on the location of the sonic point in the cooling flow that forms -- only if the sonic radius is within the galaxy scale is hot accretion possible. 
This condition was only briefly mentioned in classical studies of the cooling flow solution (\citealt{MathewsBregman78}) since in cluster-scale halos the expected
sonic radius is well within the central galaxy and thus hot mode accretion is always possible. 
We further show that our formalism for identifying the onset of hot mode accretion yields similar numerical values to the formalism in BD03, though it provides alternative physical intuition for the transition between the two regimes for galaxy accretion. Specifically, the cooling flow formalism suggests that near the threshold for hot mode accretion the halo may assume an `inverted' configuration, in which the volume-filling phase is hot and pressure-supported on large scales but cool and free-falling near the galaxy. 

To account for the possibility of a gas mass $<\fb\Mhalo$ due to e.g.~galaxy outflows, we treat the hot gas mass in our analysis as a free parameter. Our derivation thus yields for any $\Mhalo$ and redshift $z$ a maximum gas mass in which hot mode accretion is possible, or equivalently a maximum hot mode accretion rate $\Mdotcrit$ (see below).
We then compare the derived $\Mdotcrit$ to the average star formation rate (SFR) in dark matter halos at $0<z<10$, which has been constrained via abundance matching techniques and `empirical models' for how galaxies populate dark matter halos (e.g., \citealt{Moster+10, Moster+18, Behroozi+13, Behroozi+18}).

This paper is organized as follows. 
In section~\ref{s:derivation} we derive the maximum hot mode accretion rate $\Mdotcrit$ using analytic arguments, and corroborate our conclusions with idealized hydrodynamic simulations. 
In section~\ref{s:explore} we explore the dependence of $\Mdotcrit$ on halo and gas parameters, while in section~\ref{s:SFR} we compare $\Mdotcrit$ with the mean SFR in halos derived by empirical models. We summarize and discuss our results in section~\ref{s:discussion}. 
In a follow up paper (hereafter Paper III) we compare our results to the properties of halo gas in the FIRE cosmological simulations (\citealt{Hopkins+18}). 
Throughout the paper we assume a flat $\Lambda$CDM cosmology with Hubble constant $H_0=68\kms\Mpc^{-1}$ and $\Omo=0.31$ (\citealt{Planck16}).

\section{Hot vs.~cold accretion according to cooling flow solutions}\label{s:derivation}

In this section we use cooling flow solutions to derive a necessary condition for hot mode accretion, and show that this condition can be cast as a maximum hot mode accretion rate $\Mdotcrit$. In our derivation we assume that the background potential is constant in time, and limit the effects of feedback in our analysis to the possible enrichment and depletion of the halo gas, i.e.~ongoing feedback heating is assumed to be negligible. The validity of these assumptions is discussed below and tested in Paper III using cosmological simulations.

We first demonstrate in section~\ref{s:physical principles} how $\Mdotcrit$ arises by requiring $T\approx\Tvir$ and $\tcool\gtrsim\tff$ in a steady spherical flow. We then corroborate our derivation using the family of cooling flow solutions to the steady-state flow equations (section~\ref{s:stability}), and using idealized 3D hydrodynamic simulations (section~\ref{s:hydro}). 

\subsection{The $\tcool\gtrsim\tff$ condition}\label{s:physical principles}
\newcommand{\nHmax}{n_{\rm H,~max}}
\newcommand{\Mdotmax}{\dot{M}_{\rm max}}
\newcommand{\Lcool}{L_{\rm cool}}
\newcommand{\Rglx}{R_{\star}}

The energy conservation equation for a steady spherical flow is (appendix~\ref{a:Bernoulli}):
\begin{equation}\label{e:acc vs. cool}
 v_r\frac{\der}{\der r}\left(\frac{1}{2}v_r^2+\gamma\epsilon+\Phi\right) = -q ~,
\end{equation}
where $r$ is the radius, $v_r$ is the radial velocity, the sum in the brackets is the Bernoulli parameter, $\epsilon$ is the specific thermal energy, $\gamma=5/3$ is the adiabatic index, $\Phi$ is the gravitational potential, and $q$ is the cooling rate per unit mass. 
In a pressure-supported flow the kinetic term is small, while the roughly isothermal potential in dark matter halos implies that the temperature is approximately constant, so the first two terms in the brackets can be neglected. We thus get
\begin{equation}\label{e:dPhi dr}
 \frac{\der\Phi}{\der r} \approx -\frac{q}{v_r} = -\frac{\nH^2\Lambda}{\rho v_r}~,
\end{equation}
where in the second equality we replaced $q$ with $\nH^2\Lambda/\rho$ ($\rho$ and $\nH$ are the mass and hydrogen density and $\Lambda$ is the cooling function). The accretion rate $\Mdot=-4\pi r^2\rho v_r$ hence equals
\begin{equation}\label{e:Mdot}
 \Mdot \approx \frac{4\pi r^2 \nH^2\Lambda}{\der\Phi/\der r} ~. 
\end{equation}
The maximum accretion rate for the hot gas can be derived from eqn.~(\ref{e:Mdot}) by requiring that the density is low enough so $\tcool\gtrsim0.7\tff$. The motivation for the $0.7$ prefactor is given in section~\ref{s:stability}. We use 
\begin{equation}\label{e:tff}
 \tff = \frac{\sqrt{2}r}{\vc} 
\end{equation}
where $\vc$ is the circular velocity, and
\begin{equation}
\label{e:tcool}
\tcool = \frac{\epsilon}{q} = \frac{\rho\epsilon}{\nH^2 \Lambda} ~.
 \end{equation}
The maximum gas density is hence 
\begin{equation}\label{e:nHmax}
  \nHmax \approx \frac{m_{\rm p}\vc\epsilon}{0.7\cdot\sqrt{2}X \Lambda r} \approx \frac{m_{\rm p}\vc^3}{X \Lambda r} ~,
\end{equation}
where $X$ is the hydrogen mass fraction and $m_{\rm p}$ is the proton mass. In the second equality we used $\epsilon\approx\vc^2$, which is equivalent to $T\approx(4/3)(\vc/\vvir)^2\Tvir$, where $\vvir=\vc(\Rvir)$ is the virial velocity and $\Tvir=\mu m_{\rm p}\vvir^2/2k$ is the virial temperature. The prefactor in this relation is also justified in section~\ref{s:stability}. 
Plugging eqn.~(\ref{e:nHmax}) in eqn.~(\ref{e:Mdot}) and using $\der\Phi/\der r=\vc^2/r$ we get a maximum hot gas accretion rate at radius $r$ of
\begin{equation}\label{e:Mdotmax}
 \Mdotmax(r) \approx  \frac{4\pi m_{\rm p}^2\vc^4 r}{X^2\Lambda(r)} ~.
\end{equation}

In a dark matter halo with an NFW profile (\citealt{Navarro+97}) $\vc$ is roughly independent of radius, while the roughly constant temperature suggests $\Lambda$ is also approximately constant, or decreases outwards if metallicity gradients are significant. Eqn.~(\ref{e:Mdotmax}) thus suggests that $\Mdotmax(r)$ increases with radius. 
This expected increase of $\Mdotmax$ outward is robust to changes of the potential due to an average central galaxy, which is expected to cause $\vc$ near the center to fall off no faster than $\sim r^{-0.1}$ (Paper I, see figure~1 there).
Thus, for the flow to be pressure-supported at all radii we need to evaluate $\Mdotmax$ at the innermost radius of the flow. For this inner radius we use the circularization radius $\Rcirc$ at which the centrifugal and gravitational forces balance, and thus the gas can be supported by rotation rather than by thermal pressure (see further discussion below). 
The choice of $\Rcirc$ for the innermost radius of the flow is also motivated by observations which suggest galaxy sizes are $\sim\Rcirc$ (e.g., \citealt{Kravtsov13, Shibuya+15}). Using $\Rcirc$ in eqn.~(\ref{e:Mdotmax}) hence implies a maximum accretion rate for pressure-supported flows of 
\begin{eqnarray}
 \Mdotcrit &=& \Mdotmax(\Rcirc) \approx \frac{4\pi m_{\rm p}^2\vc^4\Rcirc}{X^2\Lambda} ~.\nonumber \\
           &=& 1.7 \left(\frac{\vc(\Rcirc)}{100\kms}\right)^4 \left(\frac{\Rcirc}{10\kpc}\right) \left(\frac{X^2\Lambda(\Rcirc)}{10^{-22}\,{\rm c.g.s}}\right)^{-1}\msun\yr^{-1} ~. \nonumber \\
           \label{e:Mdotcrit}
\end{eqnarray}
The numerical values of $\vc$ and $\Rcirc$ in eqn.~(\ref{e:Mdotcrit}) correspond roughly to a halo mass of $\sim4\cdot10^{11}\msun$ at $z=0$, though note that the derivation is general and applies to halos of all masses and redshifts. We estimate $\Rcirc$ using the relation
\begin{equation}\label{e:Rcirc}
 \vc(\Rcirc)\Rcirc = f_{\lambda}\frac{J_{\rm halo}}{\Mhalo} =  \sqrt{2}f_\lambda\lambda  \vvir\Rvir ~,
\end{equation}
where $J_{\rm halo}$ and $\Rvir$ are the angular momentum and virial radius of the dark matter halo, $f_\lambda$ accounts for differences between the specific angular momentum of the baryons and the average of the halo, and $\lambda$ is the halo spin parameter defined in \cite{Bullock+01}:
\begin{equation}
\lambda \equiv \frac{J}{\sqrt{2}\Mhalo \vvir\Rvir} ~.
\end{equation}
For $f_\lambda\approx1$, $\Rvir\approx200\kpc$, $\vc(\Rcirc)\approx\vvir$, and $\lambda\approx0.035$ (e.g., in the Bolshoi-Planck simulation, \citealt{RodriguezPuebla+16}) we get $\Rcirc\approx10\kpc$. Eqn.~(\ref{e:Mdotcrit}) can be further elaborated by approximating $\Lambda$ in the metal-dominated regime as (e.g.~\citealt{Wiersma+09}) 
\begin{equation}\label{e:Lambda by T} 
 \Lambda = 0.5\cdot 10^{-22} \left(\frac{T}{10^6\K}\right)^{-0.7} \left(\frac{Z}{0.3\zsun}\right)^{0.9} \erg\cm^3\s^{-1} ~,
\end{equation}
where $Z$ is the gas metallicity, and this approximation is valid at $T \sim 10^5 - 10^7\K$ and $Z\gtrsim0.3\zsun$. 
For $T=5\cdot10^5(\vc/100\kms)^2\K$ implied by $\epsilon=\vc^2$ we get
\begin{equation}\label{e:Mdotcrit num sec2}
\Mdotcrit = 0.7\left(\frac{\vc(\Rcirc)}{100\kms}\right)^{5.4} \left(\frac{\Rcirc}{10\kpc}\right)\left(\frac{Z(\Rcirc)}{0.3\zsun}\right)^{-0.9}\msun\yr^{-1} ~.
\end{equation}

\subsection{Spherically-symmetric cooling flow solutions}\label{s:stability}

To further demonstrate that hot mode accretion is possible only for accretion rates below the critical value $\Mdotcrit$ derived in the previous section, we utilize the family of cooling flow solutions derived in Paper I. 
We start by discussing purely radial flows, and then include the effects of angular momentum.

\subsubsection{Cooling flows without angular momentum}

Cooling flow solutions are derived from  the spherical steady-state equations for radiatively cooling gas in a constant gravitational potential:
\begin{eqnarray}\label{e:mass1}
& \Mdot = -4\pi r^2 \rho \vr \\
\label{e:momentum1}
& \frac{1}{2}\frac{\der \vr^2}{\der r} = -\frac{1}{\rho}\frac{\der P}{\der r} - \frac{\vc^2}{r} \\
& \vr\frac{\der \ln K}{\der r} = -\frac{1}{\tcool}~,
\label{e:energy1}
\end{eqnarray}
where $P$ is the gas pressure and $\ln K\equiv \ln (kT/\nH^{2/3})$ is the entropy. We integrate these equations as described in Paper~I, requiring the solutions to go through a sonic point and to be marginally-bound at large radii (Bernoulli parameter $\Bernoulli\rightarrow 0^-$ as $r\rightarrow \infty$). The transonic condition is required since non-transonic solutions are either not well-defined at all radii (e.g.~\citealt{Bertschinger89}), or everywhere supersonic. The exact choice of the outer boundary condition does not affect the conditions near $\Rcirc$ and hence is of no consequence for the discussion here (see Fig.~B1 in Paper I). 
For a given cooling function and gravitational potential the transonic and marginally-bound conditions yield a single-parameter family of solutions. We showed in Paper~I that gaseous halos which are initially hydrostatic converge onto these solutions within a cooling time.

\begin{figure}
\includegraphics{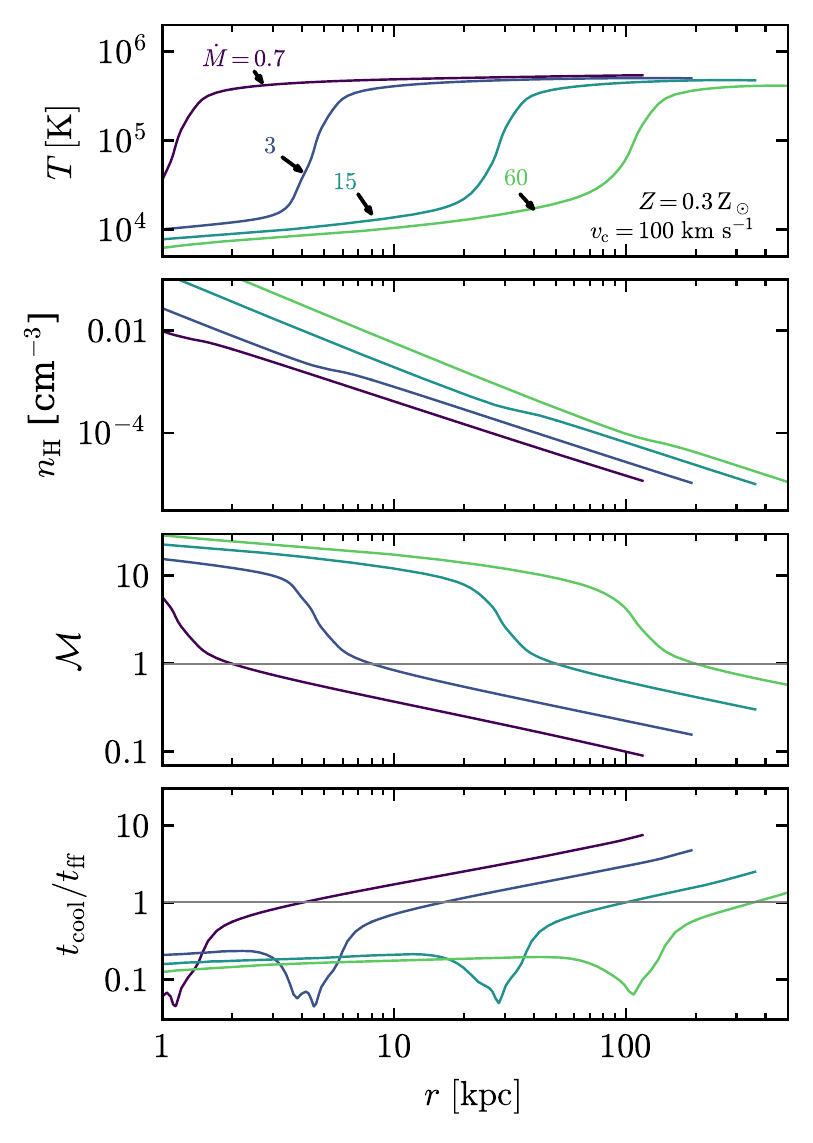}
\caption{Spherically-symmetric cooling flow solutions for the volume-filling gas phase in an isothermal potential with $\vc=100\kms$. 
The panels show the temperature, density, Mach number, and $\tcool/\tff$ of the solutions. The four solutions are derived assuming no angular momentum and $Z=0.3\zsun$, and differ in the assumed density normalization. 
The solutions are transonic, forming a cool supersonic flow with $\tcool<\tff$ within the sonic radius. A higher density normalization corresponds to a higher $\Mdot$ (noted in $\msun\yr^{-1}$ in the top panel) and to a larger sonic radius. 
}
\label{f:steady-state no AM}
\end{figure}

For simplicity we assume an isothermal gravitational potential,\footnote{To calculate the Bernoulli parameter in an isothermal potential we assume the potential equals zero at $r=10\Mpc$.} and address the implications of more realistic potentials below.  As instructive examples, we calculate four cooling flow solutions of $Z=\zsun/3$ gas in an isothermal potential with $\vc=100\kms$, corresponding at $z=0$ to $\Mhalo=4\cdot10^{11}\msun$.  
For $\Lambda$ we use the \cite{Wiersma+09} tables for $z=0$, which account for photoionization and heating by a \cite{HaardtMadau12} UV background. The panels in Figure~\ref{f:steady-state no AM} plot $T$, $\nH$, radial Mach number $\mach$, and $\tcool/\tff$ of the solutions. The solutions differ in their assumed density normalization (second panel), where a higher normalization corresponds to a higher inflow rate $\Mdot$ due to the increased cooling ($\Mdot$ indicated in the top panel) and to a larger sonic radius $\Rsonic$ (third panel). To demonstrate the dependence on $\Mdot$ in this Figure we treat the density normalization as a free parameter, though for realistic halos it is bounded from above by the halo baryon budget (see below). 

\begin{figure}
\includegraphics{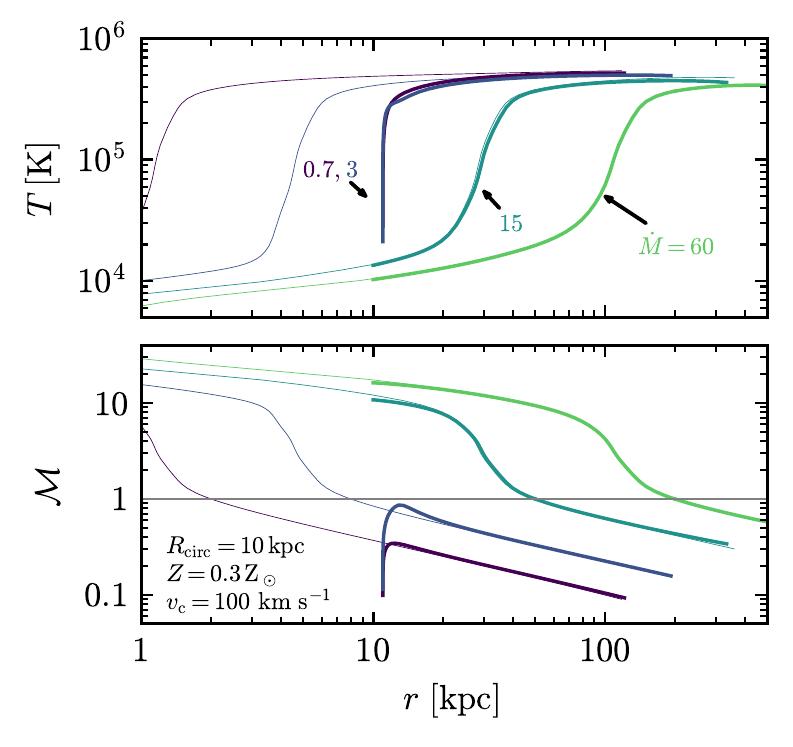}
\caption{Spherically-symmetric cooling flow solutions for gas in an isothermal potential with $\vc=100\kms$. Thick lines plot solutions for a flow with uniform specific angular momentum, corresponding to rotational support at an assumed circularization radius of $\Rcirc=10\kpc$. The values of $\Mdot$ are the same as in the no angular momentum solutions in Fig.~\ref{f:steady-state no AM} (plotted here as thin lines).
In the blue and purple solutions the flow cools just outside $\Rcirc$ and reaches $\Rcirc$  with a vanishing radial velocity. In the green and yellow solutions the flow cools at $\Rsonic>\Rcirc$, and reaches $\Rcirc$ supersonically. Solutions corresponding to hot mode accretion throughtout the halo are possible only if $\Rsonic<\Rcirc$ (where $\Rsonic$ is calculated in the no-angular momentum limit), or equivalently if $\Mdot<\Mdotcrit$ (eqn.~\ref{e:Mdotcrit}). 
}
\label{f:steady-state AM}
\end{figure}

\begin{figure*}
\includegraphics{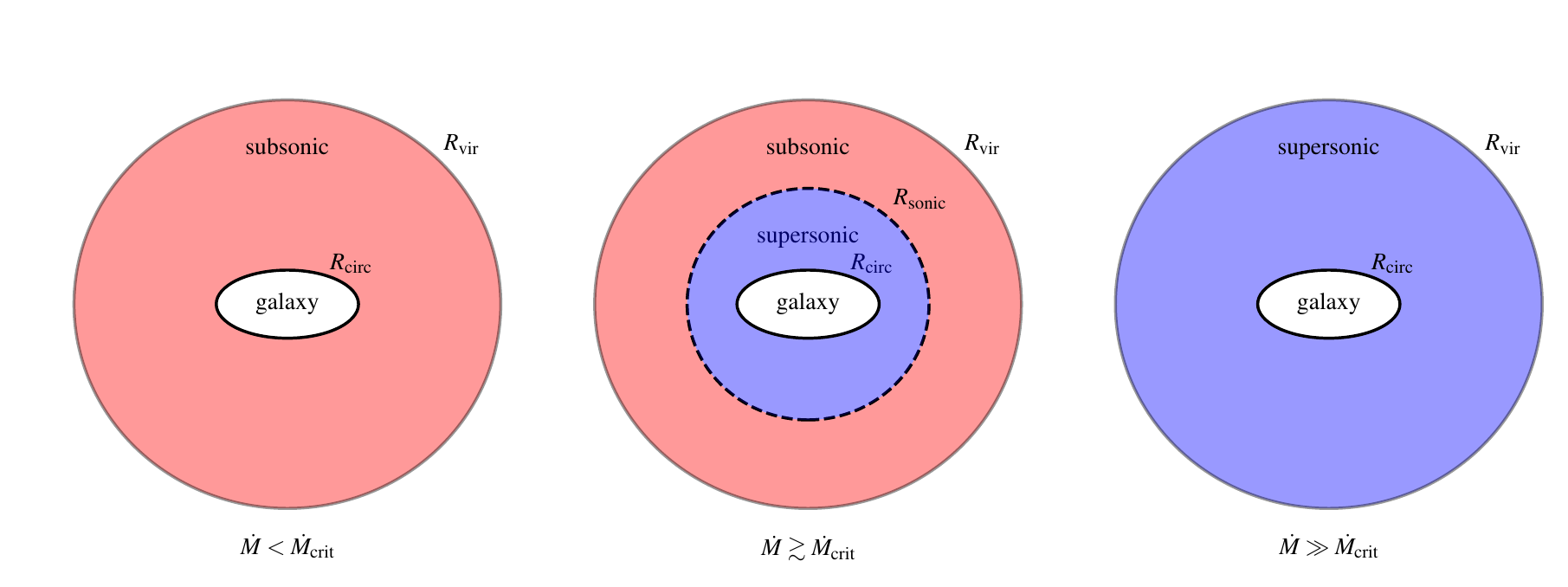}
\caption{A cartoon picturing the three types of cooling flow solutions discussed in section~\ref{s:stability} and shown in Fig.~\ref{f:steady-state AM}.  (\textbf{Left}) When $\Mdot<\Mdotcrit$ the flow is subsonic and hot ($T\approx\Tvir$) down to the circularization radius. 
(\textbf{Middle}) When $\Mdot\gtrsim\Mdotcrit$ the flow goes through a sonic point on the halo scale, and reaches the galaxy as a cool  ($T\approx10^4\K$) supersonic flow. 
(\textbf{Right}) When $\Mdot\gg\Mdotcrit$ the flow is supersonic and free falling at all halo radii. We consider only the scenario depicted in the left panel as `hot mode accretion'.} 
\label{f:cartoon}
\end{figure*}

Figure~\ref{f:steady-state no AM} shows that in the outer subsonic part of the flows the solutions satisfy the conditions for pressure support discussed in the previous section: the gas temperature is roughly equal to $\Tvir=3.6\cdot10^5\K$ (top panel) and the ratio $\tcool/\tff$ is comparable or larger than unity (bottom panel). In this region radiative cooling is balanced via heating by compression as the gas flows inward. This subsonic region can be approximated by the following self-similar solutions\footnote{These solutions correspond to the $m=0$ solutions in Paper I, where $m$ is defined such that $\vc(r)\propto r^m$.} to the flow equations~(\ref{e:mass1})--(\ref{e:energy1}), which are derived in the subsonic limit ($\mach^2\ll1$):
\begin{eqnarray}
\label{e:self similar T}
 \epsilon = \frac{9}{10}\cs^2 &=& \vc^2\\ 
\label{e:self similar nH}
\nH ~&=& ~\sqrt{\frac{\Mdot\vc^2}{4\pi \Lambda}} r^{-3/2}  \\
\label{e:self similar v}
 |\vr|~ = ~\frac{r}{\tcool} ~&=&  \frac{X}{\mpro}~\sqrt{\frac{\Mdot \Lambda}{4\pi \vc^2}} r^{-1/2}  
\end{eqnarray}
where $\cs=\sqrt{(10/9)\epsilon}$ is the adiabatic sound speed. 
From eqns.~(\ref{e:self similar T}) and (\ref{e:self similar v}), the Mach number in the self-similar solution is equal to
\begin{equation}\label{e:mach}
 \mach \equiv \frac{|\vr|}{\cs} = \frac{X}{\mpro}\sqrt{\frac{9\Mdot\Lambda}{40\pi\vc^4}} r^{-1/2} ~,
\end{equation}
i.e.~$\mach$ increases inwards as in Fig.~\ref{f:steady-state no AM}. The flow thus turns supersonic roughly at 
\begin{equation}\label{e:Rsonic}
 \Rsonic \approx \frac{9\Mdot X^2\Lambda}{40\pi\mpro^2\vc^4} ~,
\end{equation}
where this estimate is approximate due to the inaccuracy of estimating $\Rsonic$ using a solution in the subsonic limit.
The estimate for $\Rsonic$ in eqn.~(\ref{e:Rsonic}) is also an estimate of the radius where $\tcool\approx\tff$, since
\begin{equation}\label{e:tcool to tff}
 \frac{\tcool}{\tff} = \sqrt{\frac{9}{20}}\frac{\tcool}{r/\cs} = \sqrt{\frac{9}{20}}\mach^{-1} 
\end{equation}
where here we used eqns.~(\ref{e:tff}), (\ref{e:self similar T}) and (\ref{e:self similar v}). 
The sonic radius is hence roughly the radius where $\tcool/\tff=\sqrt{9/20}\approx0.7$ (see bottom panels of Fig.~\ref{f:steady-state no AM}). In section~\ref{s:physical principles} we used this prefactor and eqn.~(\ref{e:self similar T}) to derive eqn.~(\ref{e:nHmax}), though note that these relations are accurate only for an isothermal potential, and should be considered approximate in the general case.

\newcommand{\Teq}{T_{\rm eq}}

Fig.~\ref{f:steady-state no AM} shows that in the inner supersonic part of the solutions the flow rapidly loses thermal energy until it reaches the equilibrium temperature $\Teq\sim10^4\K$.  At the transition $\tcool$ is too short to be compensated by heating due to advection as in the subsonic regime, so the temperature decreases, which due to the shape of the cooling function causes the cooling to accelerate and thus further decrease the temperature. As a result of this rapid cooling process the temperature drops by a factor of $\gtrsim10$ over merely a factor of $\approx2$ in radius. 
The sonic radius thus corresponds to where the flow transitions from being largely supported against gravity via thermal pressure to being unsupported and in free-fall. 

\subsubsection{Cooling flows with angular momentum}\label{s:steady_state_AM}

Dark matter halos and the baryons associated with them are expected to have angular momentum, due to tidal torques induced by neighbouring halos. To include the effects of angular momentum in the 1D cooling flow solutions we assume a uniform specific angular momentum equal to $\vc\Rcirc$, and modify the momentum equation to (see e.g.~\citealt{Cowie+80}; BD03):
\begin{equation}\label{e:momentum AM}
\frac{1}{2}\frac{\der \vr^2}{\der r} = -\frac{1}{\rho}\frac{\der P}{\der r} - \frac{\vc^2}{r}\left[1-\left(\frac{\Rcirc}{r}\right)^2\right] ~.
\end{equation}
This equation applies to a flow within the plane defined by the angular momemtum vector. 
To derive solutions relevant for hot mode accretion, we search for solutions to the modified flow equations which satisfy $\vr\rightarrow 0$ as $r\rightarrow \Rcirc$, i.e.~the flow stalls at the circularization radius\footnote{In practice, we integrate outward from $R_0=\Rcirc(1+\epsilon_1)$ assuming $\vr(R_0)=\epsilon_2\vc$, with $\epsilon_1=\epsilon_2=0.03$. 
}. 
These solutions correspond to a radial inflow at $r\gg\Rcirc$ supported by thermal pressure which connects to a rotating flow at $r\sim\Rcirc$ supported by angular momentum. 
As an instructive example we assume $\Rcirc=10\kpc$, corresponding to $\Mhalo=4\cdot10^{11}\msun$ at $z=0$ (eqn.~\ref{e:Mdotcrit}).
We impose the same marginally-bound outer boundary condition as used for the transonic solutions discussed in the previous section. 
The thick blue and purple lines in Figure~\ref{f:steady-state AM} plot two such solutions, for $\Mdot$ equal to $0.7$ and $3\msun\yr^{-1}$ as in the corresponding non-rotating solutions from Fig.~\ref{f:steady-state no AM} (also plotted in Fig.~\ref{f:steady-state AM} as thin lines). The rotating and non-rotating solutions differ significantly only at $r\lesssim1.5\Rcirc$, where the rotating solutions stall while the non-rotating solutions continue to accelerate inward. 

For the higher values of $\Mdot$ of $15$ and $60\msun\yr^{-1}$ corresponding to the green and yellow solutions, no solutions which stall at $\Rcirc$ are possible. A transonic solution with a specific $\Mdot$ is fully-defined, and thus cannot be made to satisfy a specific boundary condition at $\Rcirc$, as is possible for the blue and purple solutions which are subsonic at all $r>\Rcirc$. The thick green and yellow lines in Fig.~\ref{f:steady-state AM} plot the corresponding transonic solution when angular momentum is included in the momentum equations. These solutions are almost identical to the no-angular momentum solutions down to $\Rcirc$, and indicate that even when angular momentum is included the flow reaches $\Rcirc$ with supersonic speeds (where it would presumably shock in a more realistic calculation). 

\begin{figure*}
 \includegraphics{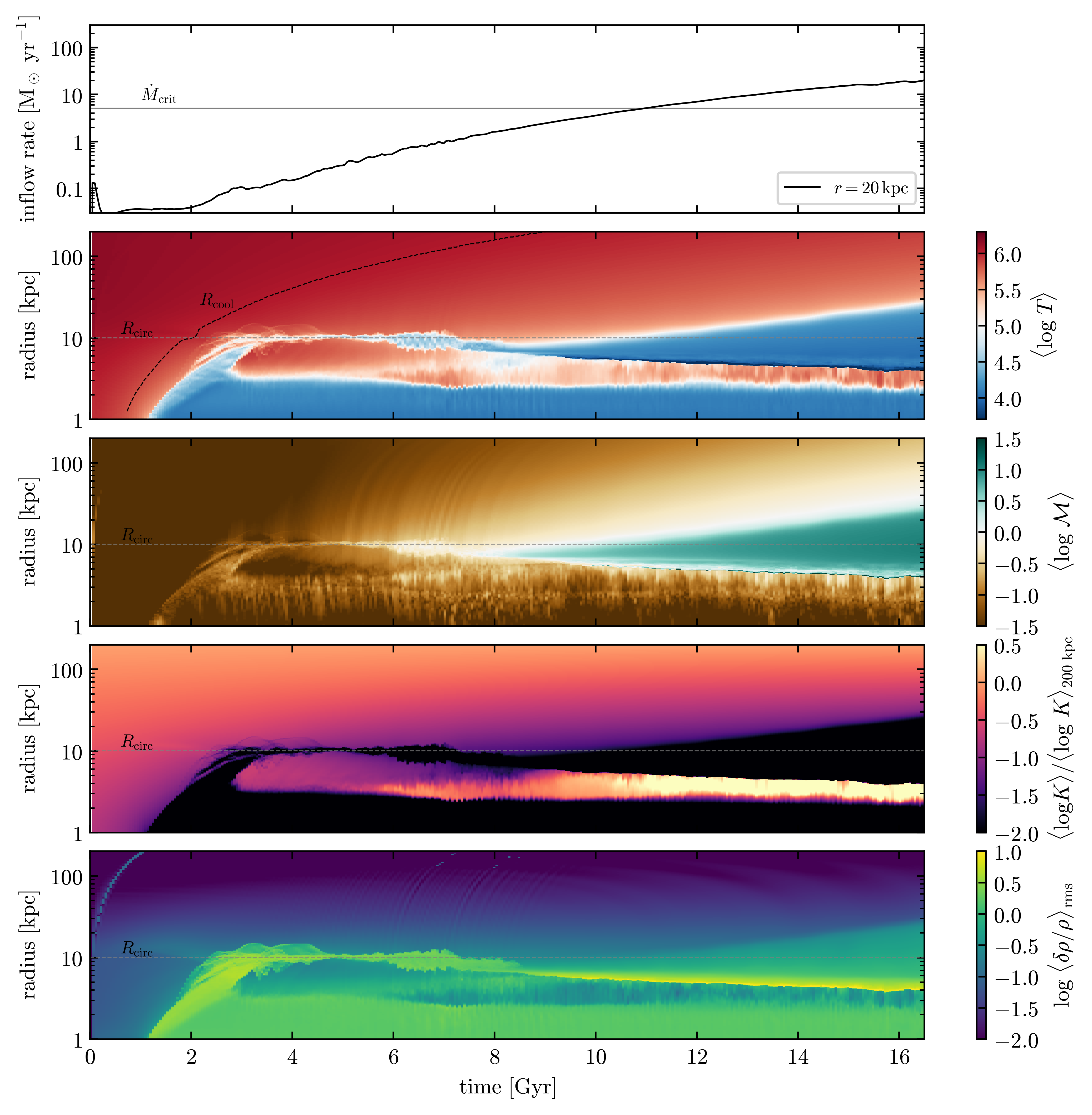}
\caption{Hydrodynamic 3D simulation of radiatively-cooling gas in an isothermal potential with $\vc=100\kms$. Third-solar metallicity is assumed throughout. The gas is initialized at $t=0$ with a hydrostatic pressure profile, and a density profile 
chosen to produce the dependence of inflow rate on time shown in the top panel. 
Initial conditions also include a uniform specific angular momentum corresponding to $\Rcirc = 10\kpc$, and density fluctations with amplitude $\sigrho = 0.03$. The three middle panels plot mass-weighted shell averages of temperature, Mach number and entropy, as a function of radius and time. Entropy is normalized by the value at $200\kpc$ at each time. The bottom panel plots the density dispersion in the shells. The critical inflow rate $\Mdotcrit$ and cooling radius $\Rcool$ are noted in the top two panels.
At $3\lesssim t \lesssim 7 \Gyr$ when $\Rcool>\Rcirc$ and $\Mdot\ll\Mdotcrit$ the halo gas forms a subsonic cooling flow corresponding to hot mode accretion -- the inward flow remains hot down to $\Rcirc$. At $t > 10 \Gyr$ when $\Mdot>\Mdotcrit$ the halo gas forms a transonic flow, with hot subsonic gas overlying cool supersonic gas. 
Density perturbations develop significantly only in supersonic regions or at $r\lesssim\Rcirc$.
}
\label{f:sim meshes}
\end{figure*}

The conclusion from Fig.~\ref{f:steady-state AM} is that only if the condition
\begin{equation}\label{e:stability criterion AM}
\Rsonic\lesssim\Rcirc
\end{equation}
is satisfied, where $\Rsonic$ is calculated in the no-angular momentum limit, then the flow can reach $\Rcirc$ with $T\approx\Tvir$ and a vanishing radial velocity. If condition~(\ref{e:stability criterion AM}) is violated as in the green and yellow solutions, than the flow necessarily reaches $\Rcirc$ supersonically. 

Figure~\ref{f:cartoon} depicts the three types of solutions discussed in this section. The left panel pictures a cooling flow with $\Mdot<\Mdotcrit$, where without angular momentum the sonic radius would be within $\Rcirc$. In this regime the flow is pressure supported on all halo scales, i.e.~the flow is subsonic and has $T\approx\Tvir$, down to the radius where the flow is supported by angular momentum. This type of flow corresponds to the classic `hot accretion mode'. 
The right panel plots solutions with $\Mdot$ sufficiently large such that the sonic radius is beyond the virial radius and hence potentially outside the accretion shock -- the outer boundary of the region in which the cooling flow solutions could be valid, since beyond the accretion shock we expect a supersonic flow. This regime corresponds to the classic cold flow regime where gas accreting from the IGM free-falls all the way down to the galaxy. The middle panel plots cooling flow solutions with $\Rcirc<\Rsonic<\rvir$, i.e~the sonic radius is within the range of radii where the cooling flow solutions could be valid. In this regime the gas is hot and pressure-supported in the outer halo, but gas in the inner halo and specifically the gas accreting onto the galaxy is cool and free-falling. In an isothermal potential this scenario applies if $1<\Mdot/\Mdotcrit < \rvir/\Rcirc\approx20$, since the sonic radius scales linearly with $\Mdot$ (eqn.~\ref{e:Rsonic}). However, since $\Rsonic\propto\vc^{-4}$, even a weak decrease of $\vc$ with increasing radius would imply that $\Rsonic$ reaches $\rvir$ at inflow rates smaller than $20\Mdotcrit$, and this intermediate regime would be relevant only over a smaller range of $\Mdot$. 

Due to the similarity of $\mach^{-1}$ and $\tcool/\tff$ in cooling flows (eqn.~\ref{e:tcool to tff}), the condition~(\ref{e:stability criterion AM}) is equivalent to the condition $\tcool\gtrsim\tff$ at $\Rcirc$ used in section~\ref{s:physical principles}. Using eqn. ~(\ref{e:stability criterion AM}) in eqn.~(\ref{e:Rsonic}) yields the maximum accretion rate of the hot mode $\Mdotcrit$, which is given by eqn.~(\ref{e:Mdotcrit}). 
We note also that \cite{QuataertNarayan00} previously discussed the importance of the sonic radius in cooling flows in isothermal potentials, and the associated mass inflow rate,  in the context of the interstellar medium of elliptical galaxies.

\subsection{Hydrodynamic simulations}\label{s:hydro}

\begin{figure}
 \includegraphics{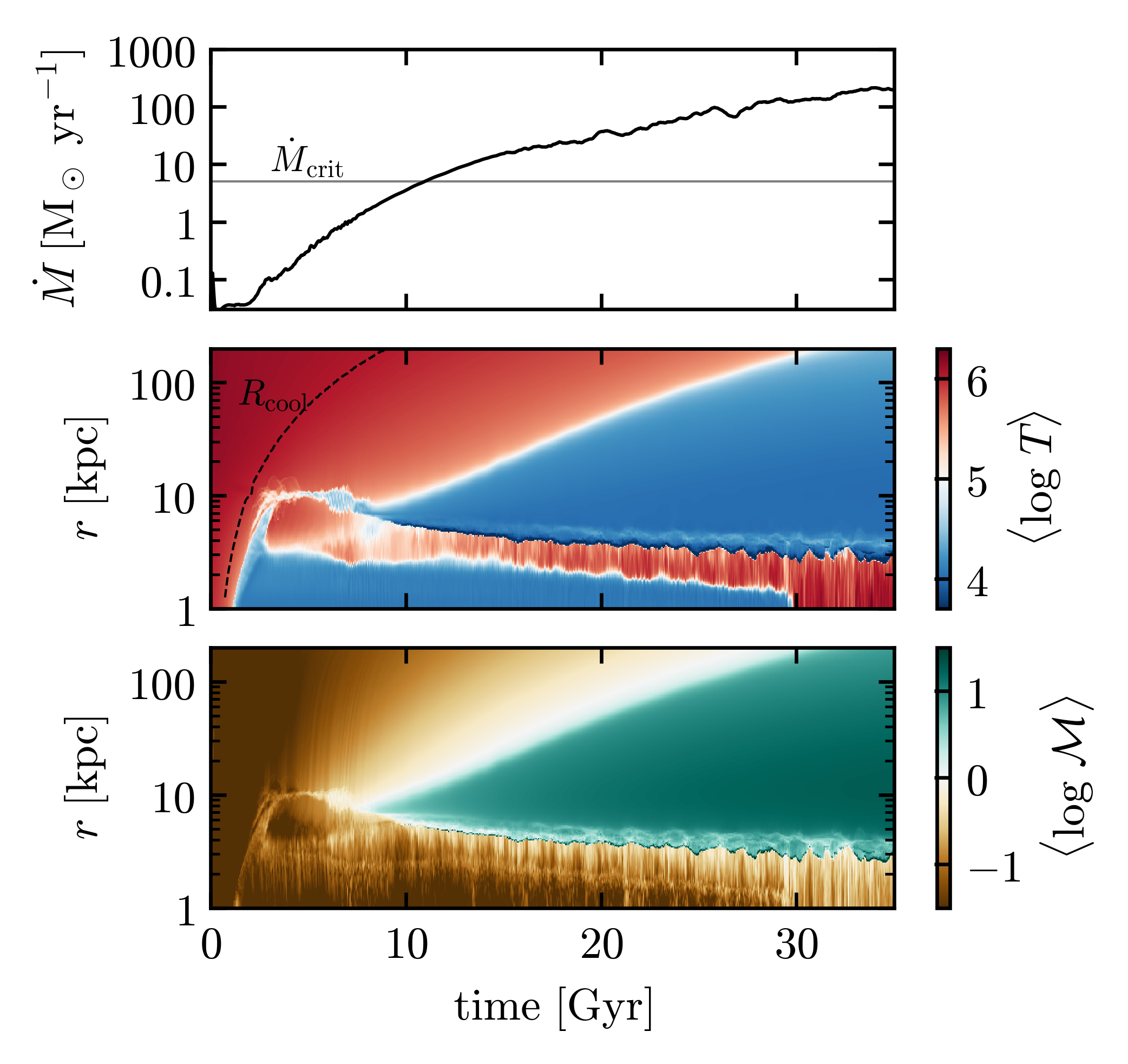}
\caption{Similar to Figure~\ref{f:sim meshes}, spanning the entire simulation time. As $\Mdot$ increases the sonic radius moves outwards. At late times when $\Mdot$ is sufficiently large, cool gas falls freely throughout the halo, corresponding to the classic `cold mode accretion'.  }
\label{f:sim meshes2}
\end{figure}

To support the above analytic results, in this section we utilize idealized 3D hydrodynamic simulations similar to the simulations used in Paper I, which are based on \cite{Fielding+17}. 
The simulations are run using the grid-based hydrodynamics code \textsc{athena++} (Stone et al., submitted\footnote{https://princetonuniversity.github.io/athena/index.html}) in a spherical-polar coordinate system. The computational domain spans $r=1\kpc-10\Mpc$, $\pi/4 \leq \theta \leq 3 \pi / 4$, and $\pi/4 \leq \phi \leq 3 \pi / 4$, where $\theta$ and $\phi$ are the polar and azimuthal angles. The grid has $64$ cells in each angular direction and $384$ logarithmically-spaced cells in $r$, which give approximately 1:1 cell aspect ratios. We adopt periodic boundary conditions in the polar and azimuthal directions, while in the radial direction we adopt outflow boundary conditions. 

We solve the standard hydrodynamics equations with additional source terms to include a static gravitational potential and radiative cooling, using the same cooling function and constant $\vc=100\kms$ as used to derive the steady-state solutions in Figures~\ref{f:steady-state no AM} -- \ref{f:steady-state AM}. 
Self-gravity of the gas is neglected.  
Angular momentum is implemented by initializing all fluid cells with a finite velocity in the $\phi$ direction $\vphi$ such that all cells outside $\Rcirc$ have the same specific angular momentum, corresponding to $\Rcirc=10\kpc$ as in Fig.~\ref{f:steady-state AM}. 
Within $\Rcirc$ we assume $v_\phi=\vc$ in the initial conditions. To avoid the accumulation of $\sim10^4\K$ gas at $\lesssim\Rcirc$ during the simulation we implement `star formation' by removing gas that satisfies $T < 3\cdot10^4\K$ and $\nH>0.03\cm^{-3}$. Tests indicate that the exact parameters of this prescription do not affect the results except where noted below.

In order to simulate the different cooling flow regimes depicted in Fig.~\ref{f:cartoon} we run a simulation where $\Mdot$ increases with time, from $\Mdot\ll\Mdotcrit$ to $\Mdot\gg\Mdotcrit$. To achieve this goal, the gas is initialized with a hydrostatic pressure profile at all radii and is allowed to radiatively cool. As demonstrated in Paper I from this initial configuration the flow is expected to converge onto one of the steady-state cooling flow solutions, at radii smaller than the cooling radius (see figures 6--9 there). The mass inflow rate in the cooling flow that forms is expected to evolve as (\citealt{Bertschinger89}, hereafter B89): 
\begin{equation}\label{e:Mdot stable}
 \Mdot_{\rm B89}(t,r\ll\Rcool) \approx 4\pi \Rcool^2\rho(\Rcool)\frac{\der \Rcool}{\der t} ~,
\end{equation}
where $\Rcool(t)$ is the cooling radius at which $\tcool=t$. This relation represents a `cooling wave' expanding in the initially static medium at a velocity $\der\Rcool/\der t$, as gas at increasingly larger radii starts cooling and joins the cooling flow.
For a constant initial temperature with sound speed $\cs$, hydrostatic equilibrium gives $\rho\propto r^{-\alpha}$ with $\alpha=\gamma\vc^2/\cs^2$. The cooling time thus scales as $\tcool\propto r^\alpha$ (eqn.~\ref{e:tcool}), and the cooling radius as $\Rcool\propto t^{1/\alpha}$. Equation~(\ref{e:Mdot stable}) hence yields
\begin{equation}\label{e:Mdot stable 2}
 \Mdot_{\rm B89}(t) \propto t^{3/\alpha-2} ~,
\end{equation}
where the constant of proportionality is determined by the normalization of the initial density profile. 
We choose an initial density profile\footnote{In a general hydrostatic profile the density slope $\alpha$ is a free parameter. In contrast, steady-state cooling flows in an isothermal potential have a specific density slope of $\nH\propto r^{-1.5}$ (eqn.~\ref{e:self similar nH}).} with $\alpha=0.5$ (i.e.~an initial sound speed of $\cs=2\gamma\vc^2$ and initial temperature of $1.5\cdot10^6\K$) and a hydrogen particle density $\nH=10^{-4.5}\cm^{-3}$ at $r=100\kpc$. We emphasize that this choice of initial conditions is intended to yield a desired $\Mdot(t)$ rather than to describe a realistic halo. 
We also impose in the initial conditions small isobaric density perturbations with an amplitude $\sigrho = 0.03$ and a white-noise spectrum.

The top panel of Figure~\ref{f:sim meshes} plots $\Mdot(t)$ in the main simulation, which scales roughly as $\Mdot\sim t^4$ as expected from eqn.~(\ref{e:Mdot stable 2}). The plotted $\Mdot(t)$ is measured just beyond $\Rcirc$ at $r=20\kpc$, though our results do not depend on the exact choice of radius since at a given snapshot $\Mdot$ varies by less than $30\%$ in the range $\Rcirc < r < \Rcool/3$ ($\Rcool$ is plotted in the second panel). At $t=11\Gyr$ the inflow rate exceeds $\Mdotcrit=5.1\msun\yr^{-1}$, where $\Mdotcrit$ is calculated via eqn.~(\ref{e:Mdotcrit}). 
The lower four panels of Fig.~\ref{f:sim meshes} plot shell-averaged properties in the simulation as a function of of $r$ and $t$. From top to bottom the panels show the mass-weighted averages of $\log T$, $\log\mach$, and $\log K/K(r=200\kpc)$, and the density dispersion $\sigrho$. 
The figure shows that at early times $t\lesssim2\Gyr$ the value of $\Rcool$ is smaller than $\Rcirc$ and the gas properties remain near the initial conditions. At later times $3 \lesssim t \lesssim 10\Gyr$ when $\Rcool>\Rcirc$ and $\Mdot<\Mdotcrit$, the gas within $\Rcool$ forms a subsonic cooling flow in which the gas temperature is near virial and the entropy declines inward. Only very close to $\Rcirc=10\kpc$ the flow cools out, as suggested by the 1D solutions with $\Mdot=0.7$ and $3\msun\yr^{-1}$ in Fig.~\ref{f:steady-state AM}. Comparing snapshots in the simulation with a steady-state solution with the same $\Mdot$ as in the snapshot, we find that the mass-weighted $T$, $\rho$ and $v$ differ by a factor of less than two at $\Rcirc<r<\Rcool/3$, and a factor of less than $1.5$ at $\Rcirc<r<\Rcool/5$, consistent with the result in Paper I that the flow converges onto the steady-state solutions at $r\ll\Rcool$, and specifically near $\Rcirc$. Fig.~\ref{f:sim meshes} shows also that within $\Rcirc$ the flow is hot down to $\approx 3\kpc$, and comprises of inflow outside the midplane which overshoots $\Rcirc$ and is then repelled back by the centrifugal force (see Fig.~\ref{f:flowlines} below). The lowest panel demonstrates that in this $\Mdot<\Mdotcrit$ regime the amplitude of density fluctations is $\ll1$ beyond $\Rcirc$, as expected in subsonic cooling flows (see Paper I and references therein).

\begin{figure}
 \includegraphics{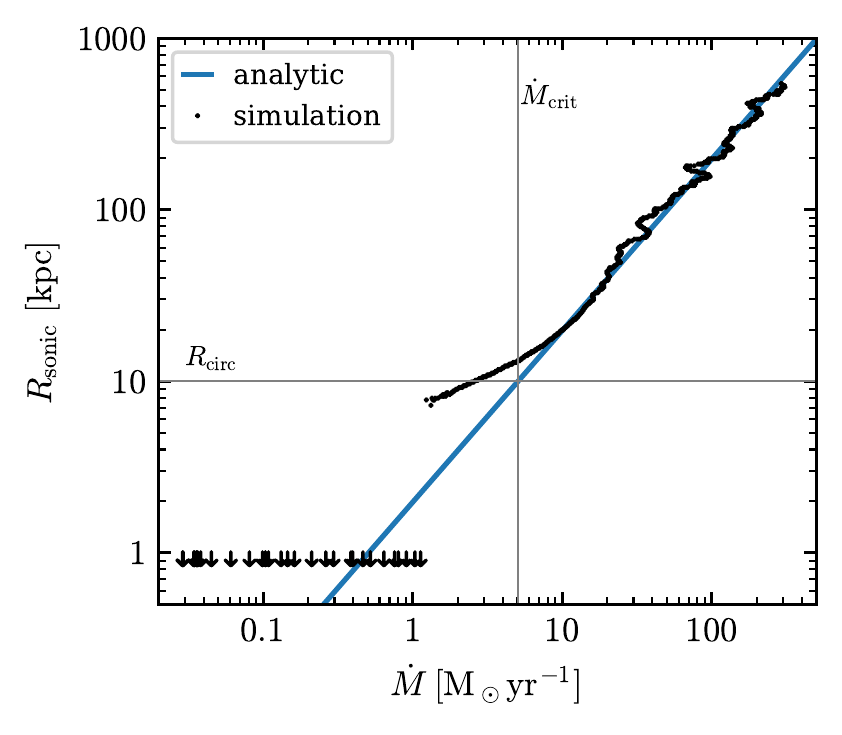}
\caption{The relation between $\Mdot$ and $\Rsonic$. Each dot or arrow corresponds to a snapshot in the simulation shown in Figs.~\ref{f:sim meshes}--\ref{f:sim meshes2}, while down-pointing arrows at $1\kpc$ denote snapshots where the flow is entirely subsonic. The cyan line plots the analytic relation (eqn.~\ref{e:Rsonic}). The simulation and analytic calculations roughly agree when $\Rsonic>\Rcirc$ and $\Mdot>\Mdotcrit$. }
\label{f:Rsonic}
\end{figure}

Fig.~\ref{f:sim meshes} shows that at $t>10\Gyr$ when $\Mdot$ exceeds $\Mdotcrit$ the flow is transonic with a subsonic region `overlying' a supersonic region, as depicted in the middle panel of Fig.~\ref{f:cartoon}. The sonic point is evident as the upper white contour in the $\mach$ panel, and it moves outward as $\Mdot$ increases in the simulation (see Figure~\ref{f:sim meshes2}). We measure $\Rsonic$ as the outermost shell with $\avgmach=1$, and plot the relation between $\Rsonic$ and $\Mdot(r=20\kpc)$ in Figure~\ref{f:Rsonic}. When $\Mdot>\Mdotcrit$, the relation in the simulation is similar to the steady-state, no-angular-momentum calculation (eqn.~\ref{e:Rsonic}, blue line). When $\Mdot\lesssim\Mdotcrit$, i.e.~at $7.5\lesssim t \lesssim 10\Gyr$, the simulation has a sonic point somewhat within $\Rcirc$, while when $\Mdot\ll\Mdotcrit$ the flow in the simulation is entirely subsonic.

Towards the end of the simulation where $\Mdot\gtrsim20\Mdotcrit$, Fig.~\ref{f:sim meshes2} shows that the sonic radius exceeds $200\kpc$, and the flow is supersonic at all halo scales, i.e.~the scenario depicted in the right panel of Fig.~\ref{f:cartoon}. We note that in our simulation the flow is transonic even at these late times since there is hot quasi-static gas out to the outer boundary at $10\Mpc$, i.e.~we effectively assume the accretion shock is at infinity. In a realistic system with a finite accretion shock radius we expect the flow to be purely supersonic if $\Mdot$ corresponds to $\Rsonic$ larger than the shock radius.

Figs.~\ref{f:sim meshes} and \ref{f:sim meshes2} demonstrate that once the flow crosses the sonic radius it cools quickly, as suggested by the steady-state solutions shown in Fig.~\ref{f:steady-state no AM}. This rapid cooling is associated with a rapid growth of thermal instabilities (bottom panel in Fig.~\ref{f:sim meshes}, see also \citealt{MathewsBregman78} and \citealt{BalbusSoker89}). The association of the sonic radius with the rapid growth of instabilites occurs since within the sonic radius $\tcool\ll r/\vr$ (see figure~3 in Paper I), so the instabilities grow faster than the rate at which the flow is advected inward, in contrast with the subsonic region where $\tcool\approx r/\vr$ (eqn.~\ref{e:self similar v}). The supersonic flow reaches a radius which is substantially smaller than $\Rcirc$, and is evident as a boundary in all properties plotted in Fig.~\ref{f:sim meshes}. This minimum radius decreases with increasing $\Mdot$.

\begin{figure}
 \includegraphics{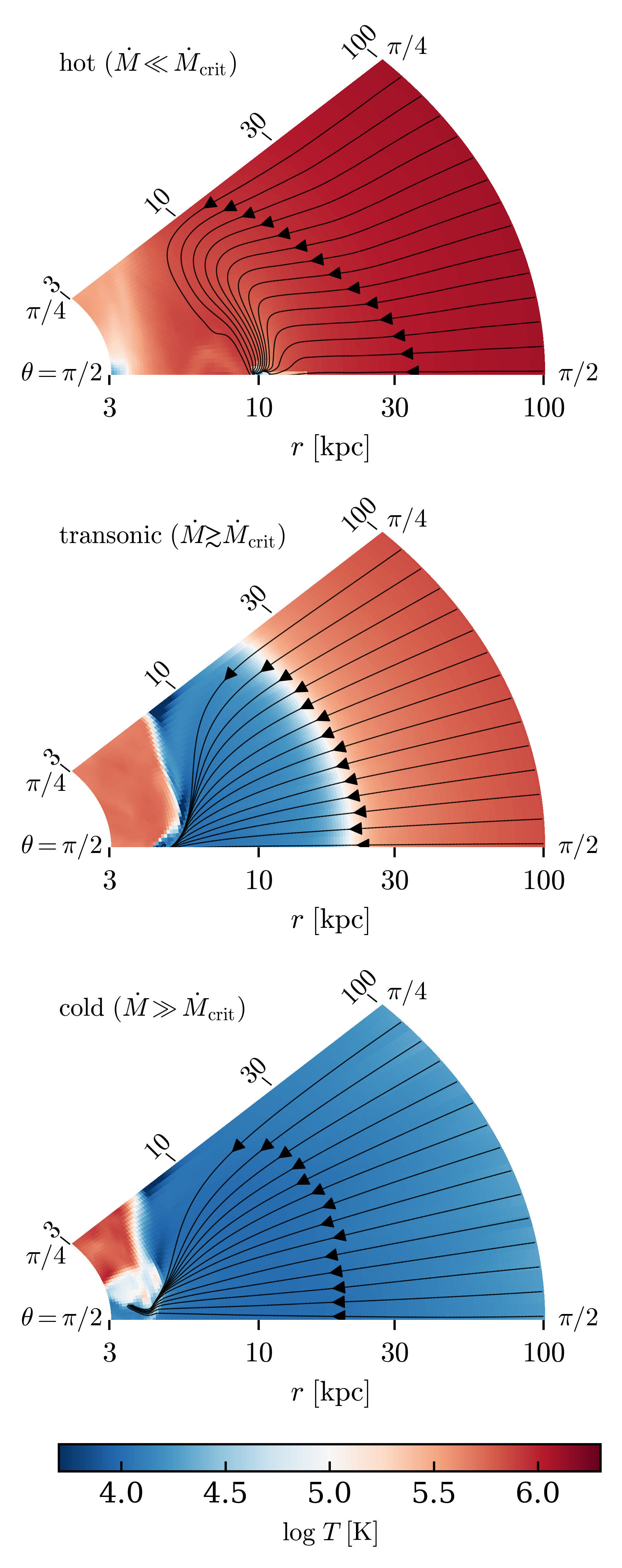}
\caption{Streamlines and temperature maps in the meridional plane. The panels plot three snapshots of the simulation, corresponding to the hot accretion mode ($t=4\Gyr$, top), the transonic accretion mode ($t=15\Gyr$, middle),  and the purely cold accretion mode ($t=30\Gyr$, bottom). The bottom axis of each panel corresponds to the midplane ($\theta=\pi/2$) while the top diagonal axis corresponds to the boundary of the simulated domain at $\theta=\pi/4$. The equilibrium position given the angular momentum of the simulated gas is at $r=\Rcirc=10\kpc$ and $\theta=\pi/2$.
In the hot accretion mode the radial flow at large scales converges onto the equilibrium position, at which point the gas cools and is removed from the simulation via our SF prescription. In the other two accretion modes the flow reaches the midplane and is lost to SF at radii smaller than the equilibrium position.
}
\label{f:flowlines}
\end{figure}

In Figure~\ref{f:flowlines} we plot streamlines and temperature maps in the meridional plane, mass-weighted over the $\phi$ coordinate. The three panels plot snapshots at $t=4\Gyr$ (top), $t=15\Gyr$ (middle) and $t=30\Gyr$ (bottom), corresponding to the hot ($\Mdot\ll\Mdotcrit$), transonic ($\Mdot\gtrsim\Mdotcrit$) and purely cold ($\Mdot\gg\Mdotcrit$) accretion phases. The panels are shaped as wedges similar to half the simulated domain, between $\theta=\pi/2$ (bottom axis) and $\theta=\pi/4$ (top diagonal axis). The streamlines emanate from large radii and are initially evenly spaced in $\theta$, indicating a radial inflow. In the hot accretion mode plotted on top the streamlines converge onto $r=\Rcirc$ and $\theta=\pi/2$, i.e.~on the equilibrium position for our assumed specific angular momentum, which corresponds to a `ring' in 3D space. Streamlines initially far away from the midplane first overshoot $\Rcirc$ and reach somewhat smaller radii, and then turn outward as the centrifugal force overcomes gravity. 
Note though that this latter effect may be artificially enhanced by the boundary of our domain at $\theta=\pi/4$ and hence the lack of streamlines which feed gas and provide pressure support at smaller radii. At the equilibrium position the gas is cool (see also temperature panel of Fig.~\ref{f:sim meshes}), though the flow cools out just before joining the ring -- the region with $T\lesssim10^5\K$ spans $\lesssim3$ grid cells in the $\theta$ direction and $\lesssim9$ grid cells in the $r$ direction.  In the ring our prescription for `star formation' acts as a sink for gas when the density exceeds $0.03\cm^{-3}$. 

In contrast with the hot accretion mode, in the other two regimes shown in the bottom panels of Fig.~\ref{f:flowlines} the streamlines reach the midplane at a radius of $4-5\kpc$, substantially smaller than the equilibrium position at $r=\Rcirc$. This is possible due to the lack of pressure support and high inertia of the flow, which allows a cool flow to `overshoot' the angular momentum barrier. In our simulation the gas is lost to SF at these inner radii, which causes the flowlines to end at a non-equilibrium position. In a similar simulation without the SF prescription the flow circles back to the equilibrium position after crossing the midplane at radii $<\Rcirc$.

To summarize, the flow structure formed in the 3D simulation suggests that the 1D steady-state cooling flow solutions capture the transition between hot and cold mode accretion reasonably well, at least in our idealized setup. If $\Mdot<\Mdotcrit$ then the flow `smoothly' accretes onto the galaxy disk from a hot ($\approx\Tvir$) rotating atmosphere with a vanishing radial velocity, as can be seen in the purple and blue solutions in Fig.~\ref{f:steady-state AM} and in the top panel of Fig.~\ref{f:flowlines}. In contrast if $\Mdot>\Mdotcrit$ then  $\Rsonic>\Rcirc$ and the gas reaches the galaxy scale as a cool ($\approx\Teq$) supersonic flow, as can be seen in the green and yellow solutions in Fig.~\ref{f:steady-state AM} and in the two bottom panels of Fig.~\ref{f:flowlines}. 

We note that our result where initially hydrostatic gas converges onto a steady-state cooling flow solution at $\ll \Rcool$ requires that $\der\Rcool/\der t < \cs$, i.e.~the cooling wave expands slowly compared to the sound-crossing time. This condition was also required by B89 in order to derive their self-similar cooling wave solutions. If this condition is violated, gas at different radii cools out monolithically, and the pressure profile does not have time to adjust to the cooling flow solution. In this latter case the halo gas collapses into a supersonic free-falling solution rather than forming a subsonic or transonic cooling flow. In appendix~\ref{a:collapse} we show that this collapse occurs in one of the simulations presented in Paper I.

\subsection{Comparison to the condition for shock stability}\label{s:BD03}
\renewcommand{\Rsh}{R_{\rm sh}}
\newcommand{\vsh}{v_{\rm sh}}

The simulation in the previous section and the simulations in Paper I demonstrate that halo gas which is initially hydrostatic converges onto the family of cooling flow solutions (as long as $\der \Rcool/\der t\ll\cs$, see appendix~\ref{a:collapse} and B89). A related question is under which conditions a flow which is initially supersonic\footnote{The family of supersonic solutions to eqns.~(\ref{e:mass1})--(\ref{e:energy1}) satisfies $T\approx\Teq\approx10^4\K$ and $\vr$ roughly equal to the free-fall velocity. As $\rho$ is essentially unconstrained, one can find such a supersonic solution for any assumed value of $\Mdot$.} will shock and form a cooling flow. Note that such a transition is non-trivial only if $\Rsonic$ of the cooling flow that forms is smaller than the outer boundary of the system (e.g.~the two left panels in Fig.~\ref{f:cartoon}), since otherwise the cooling flow solution is also a supersonic solution (right panel in Fig.~\ref{f:cartoon}). This question was addressed by BD03, who argued that supersonic inflows in dark matter halos shock once the conditions for an accretion shock to expand are met at the disk radius $\approx\Rcirc$. In this section we show that the condition for an expanding accretion shock at a shock radius $\Rsh=\Rcirc$ is similar to the condition $\Rsonic<\Rcirc$ derived here for the onset of hot mode accretion. We show this similarity by utilizing the expectation that the postshock gas forms a cooling flow\footnote{This expectation is not strictly valid, since cooling flows are steady-state solutions while gas immediately within the shock radius is likely not time-steady (as is gas just within $\Rcool$ in the simulation discussed in \S\ref{s:hydro}). The cooling flow solutions for the shocked gas are expected to be accurate only out to radii smaller than the shock radius. We neglect this complication, and in the next section support our conclusion on the similarity of the two formalisms by comparing the values they yield for $\Mthres$.}.

The shock jump condition is
\begin{equation}
 v_{\rm sh} = -\frac{1}{3}\left(v_0 - 4v_1\right) ~,
\end{equation}
where $v_{\rm sh}$, $v_0$, and $v_1$ are the shock, preshock, and postshock velocities, all measured in the halo frame (inflows have a negative velocity), and for simplicity we assume a strong shock. Note that the postshock velocity in the shock frame $v_1 - v_{\rm sh}$ must be subsonic, so if $v_1$ is supersonic $v_{\rm sh}$ must be negative, i.e.~the shock is contracting. It hence follows that a necessary condition for an expanding accretion shock is $\mach(\Rsh)<1$, or equivalently in a cooling flow $\Rsonic<\Rsh$. To show that the $\Rsonic<\Rsh$ condition is likely to be also a sufficient condition for an expaning shock, we replace $v_1$ with $-R_{\rm sh}/\tcool$ (eqn.~\ref{e:self similar v}):
\begin{equation}\label{e:v_sh0}
 v_{\rm sh} = -\frac{1}{3}\left(v_0 + 4\frac{R_{\rm sh}}{\tcool}\right) 
\end{equation}
Using the definition of $\tff$ (eqn.~\ref{e:tff}) and extracting $v_0$ from the parentheses we get
\begin{equation}\label{e:v_sh}
 v_{\rm sh} = -\frac{v_0}{3}\cdot\left(1-\frac{\tff/\tcool}{|v_0|/(\sqrt{8}\vc)}\right) ~,
\end{equation}
where all quantities are estimated at $R_{\rm sh}$. Approximating the inflow velocity as $|v_0|\approx \vc(1+\sqrt{2\ln(\Rvir/R_{\rm sh})})$, as expected for an inflow `dropped' from $2\rvir$ in an NFW potential, we get $|v_0|/\sqrt{8}\vc = 1.2$ for $R_{\rm sh}=0.05\rvir$. 
Eqn.~(\ref{e:v_sh}) thus implies that if $\tff/\tcool\lesssim1.2$ at the shock radius, then the term in the brackets is positive and the shock expands outward. Since $\tff/\tcool\approx 1.5\mach$ (eqn.~\ref{e:tcool to tff}), we get that if $\mach(\Rcirc)\lesssim0.8$ then $\vsh>0$, i.e.~if the condition $\Rsonic\lesssim\Rcirc$ is satisfied the shock would expand. 

The above derivation suggests that the BD03 condition for shock stability at $\Rcirc$ is similar to the condition for hot mode accretion $\Rsonic\lesssim\Rcirc$ derived in this work. 
It is important to note though that our derivation does not assume an accretion shock exists, in contrast with the derivation of BD03. Rather, our derivation is based solely on the properties of radiatively cooling gas with $T\approx\Tvir$, regardless of whether the gas was heated to this temperature in a single shock, in a series of shocks, or by feedback at earlier epochs. 
Our analysis thus suggests that the conditions under which hot mode accretion is possible apply more generally. 

In the simulations in BD03 the halo mass grows with time, so $\Mdot/\Mdotcrit$ decreases since $\Mdotcrit$ increases faster than $\Mdot$ (see next section). The initially supersonic flows in BD03 though shock directly into subsonic flows, without going through an intermediate transonic cooling flow phase. We have verified this behavior using a setup similar to that in the previous section but with supersonic initial conditions. We set the outer boundary condition so $\Mdot$ decreases with time from an initial $\Mdot\gg\Mdotcrit$, and indeed a shock and subsonic cooling flow form only when $\Mdot\lesssim\Mdotcrit$, while when $\Mdot\gtrsim\Mdotcrit$ the flow remains purely supersonic rather than forming a transonic flow. A possible limitation of this simulation and the simulation in BD03 is the lack of sufficiently strong shocks beyond $\Rcirc$. 
If the flow shocks in the supersonic part of the flow the shock cannot propagate outward, and hence the subsonic part of the cooling flow will not form. However, in the presence of outflows from the galaxy the inflows from the IGM are expected to experience strong shocks at radii $\gg\Rcirc$ (e.g.~\citealt{Fielding+17}), and thus supersonic flows may shock directly into transonic cooling flows. We leave exploring this possibility to future work. 

\subsection{The condition for cooling-regulated accretion}\label{s:WF91}

\cite{WhiteFrenk91} argued that the condition 
\begin{equation}\label{e:WF91}
 \tcool(\rvir)=\thubble
\end{equation}
separates between `cooling-limited' systems in which accretion is regulated by radiative cooling, and `supply-limited' systems in which accretion is regulated by the inflow rate from the IGM. 
We now show that eqn.~(\ref{e:WF91}) is similar to the condition $\tcool=\tff$ at $\Rcirc$ derived above for the onset of hot mode accretion. This similarity follows since in cooling flows $\tcool/\tff\propto r^{1/2}$ (eqn.~\ref{e:tcool to tff}), so in the critical solution $\tcool/\tff \propto (r/\Rcirc)^{1/2}$ and hence
\begin{equation}\label{e:WF91 comparison}
 \tcool(\rvir, \Mdot=\Mdotcrit) \approx \sqrt{\frac{\Rvir}{\Rcirc}}\tff(\rvir)\approx \thubble
\end{equation}
where the last approximation follows from $\Rvir/\Rcirc\approx20$ and $\thubble\approx5\tff(\rvir)$.\footnote{The relation $\thubble=5\tff(\rvir)$ can be derived from $\tff(\rvir)=\sqrt{2}\rvir/\vvir \approx \sqrt{2}/(10H)$ and $\thubble=2/(3H)$, where $H$ is the Hubble parameter.} Thus, systems with $\Mdot<\Mdotcrit$ are cooling-limited according to the condition~(\ref{e:WF91}), while systems with $\Mdot>\Mdotcrit$ are `supply-limited', even if the halo gas has shocked and forms a transonic cooling flow. 

We note in passing that semi-analytic models such as \cite{SomervillePrimack99} which employ the condition~(\ref{e:WF91}) could be improved by considering the halo gas density profile $\rho\propto r^{-1.5}$ suggested by the physical cooling flow solution (eqn.~\ref{e:self similar nH}), rather than say an isothermal profile with $\rho\propto r^{-2}$.

\section{The critical cooling rate as a function of halo and gas parameters}\label{s:explore}

We now use eqn.~(\ref{e:Mdotcrit}) to evaluate $\Mdotcrit$ as a function of halo parameters. We use the following virial relations: 
\begin{eqnarray}
\label{e:Rvir}
 \Rvir &=& \frac{\vc}{\sqrt{\frac{\Deltac(z)}{2}} H(z)} = 263\, M_{12}^{1/3} \Hnormp{-2/3}\kpc \\ 
\label{e:vc}
 \vvir &=& \left(\sqrt{\frac{\Deltac}{2}} H G \Mhalo \right)^{1/3} = 128\, M_{12}^{1/3} \Hnormp{1/3}\kms \nonumber \\
\end{eqnarray}
where $\Mhalo\equiv10^{12}M_{12}\msun$, $\Deltac$ is the virial overdensity with respect to the critical density from \cite{BryanNorman98}, $H(z)$ is the Hubble parameter at redshift $z$, and we absorbed the redshift-dependent term $\sqrt{\Deltac(z)} H(z)/\sqrt{\Deltac(0)}H_0$ into a function $\Hnorm$, which is equal to
\begin{equation}
 \Hnorm = \sqrt{\frac{\Deltac(z)}{102}\left[1-\Omo+\Omo(1+z)^3\right]}\approx (1+z)^{0.9}
\end{equation}
where the approximation is accurate to 25\%\ at $0<z<10$. 
The critical accretion rate is then derived using eqns.~(\ref{e:Rvir}), (\ref{e:vc}) and (\ref{e:Rcirc}) in eqn.~(\ref{e:Mdotcrit}): 
\begin{eqnarray}\label{e:Mdotcrit num}
 \Mdotcrit &=& \frac{2^{1/6}\cdot4\pi G^{5/3}\mpro^2}{X^2} f_{\vc}^3 f_\lambda\lambda\Deltac^{1/3} H^{2/3} \Mhalo^{5/3}\Lambda^{-1} \nonumber \\
            &=& 10.6 \, \fvc^3 f_\lambda\lambda_{0.035} \Lambda_{-22}^{-1} M_{12}^{5/3} \Hnormp{2/3}\msun\yr^{-1} ~, \nonumber\\
\end{eqnarray}
where we defined 
\begin{equation}
\fvc\equiv \frac{\vc(\Rcirc)}{\vvir},
\end{equation}
which depends both on the halo concentration and on the properties of the galaxy.
In the numerical evaluation in eqn.~(\ref{e:Mdotcrit num}) we used $X=0.7$, and defined $\Lambda\equiv10^{-22}\Lambda_{-22}\erg\cm^3\s^{-1}$ and $\lambda\equiv0.035\lambda_{0.035}$. 
The value of $\lambda$ is normalized to the mean value found in the Bolshoi-Planck simulation in halos with mass $10^{10}<\Mhalo<10^{15}\msun$ and redshift $0<z<8$ (\citealt{RodriguezPuebla+16}).

We now describe how we estimate $\Lambda$ and $\fvc$ in eqn.~(\ref{e:Mdotcrit num}). For $\Lambda$ we use the tables of \cite{Wiersma+09}, which depend on $T$, $Z$, $z$, and $\nH$. 
The value of $T$ is calculated from $\epsilon=\vc^2$ (eqn.~\ref{e:self similar T}) which gives
\begin{equation}\label{e:Tvir}
 k T = \frac{2}{3}\mu\mpro \vc^2 = \frac{4}{3}\fvc^2 k\Tvir ~.
\end{equation}
For $Z$ we use as a fiducial estimate the metallicity of gas in the central galaxy, since at $\Rcirc$ which is roughly the size of the galaxy (e.g.~\citealt{Kravtsov13, Shibuya+15}) significant mixing of the hot gas and ISM is likely. We calculate this metallicity estimate based on the observed $z=0$ mass-metallicity relation from \cite{AndrewsMartini13}:
\begin{equation}\label{e:MZR}
 \frac{\zism(\Mhalo,z=0)}{\zsun} =  \frac{1.28}{1+\left(\frac{\Mstar(\Mhalo,z=0)}{10^{8.901}\msun}\right)^{-0.64}} ~,
\end{equation}
where we converted $[{\rm O}/{\rm H}]$ in \cite{AndrewsMartini13} to $\zism/\zsun$ assuming $12 + [{\rm O}/{\rm H}]_{\odot}=8.69$ (\citealt{Asplund+09}), and we use the stellar-mass halo-mass relation (SMHM) from \citeauthor{Behroozi+18} (2019, hereafter B19) to convert between $\Mhalo$ and stellar-mass $\Mstar$. 
The dependence of $\Lambda$ on $\nH$ and $z$ is due to heating and ionization by the UV background (UVB) and cooling off the cosmic microwave background, where only the UVB effect is significant in the halo masses of interest. 
To calculate $\nH$ we solve eqn.~(\ref{e:nHmax}) for $\nHmax$ including the dependence of $\Lambda$ on $\nH$, and then rederive $\Mdotcrit$ accordingly. To gauge the importance of the UVB on $\Mdotcrit$ we also calculate $\Lambda$ assuming no UVB (i.e., in the $\nH\rightarrow\infty$ limit), using the collisional-ionization equilibrium cooling tables from \cite{GnatSternberg07}.

\begin{figure}
 \includegraphics{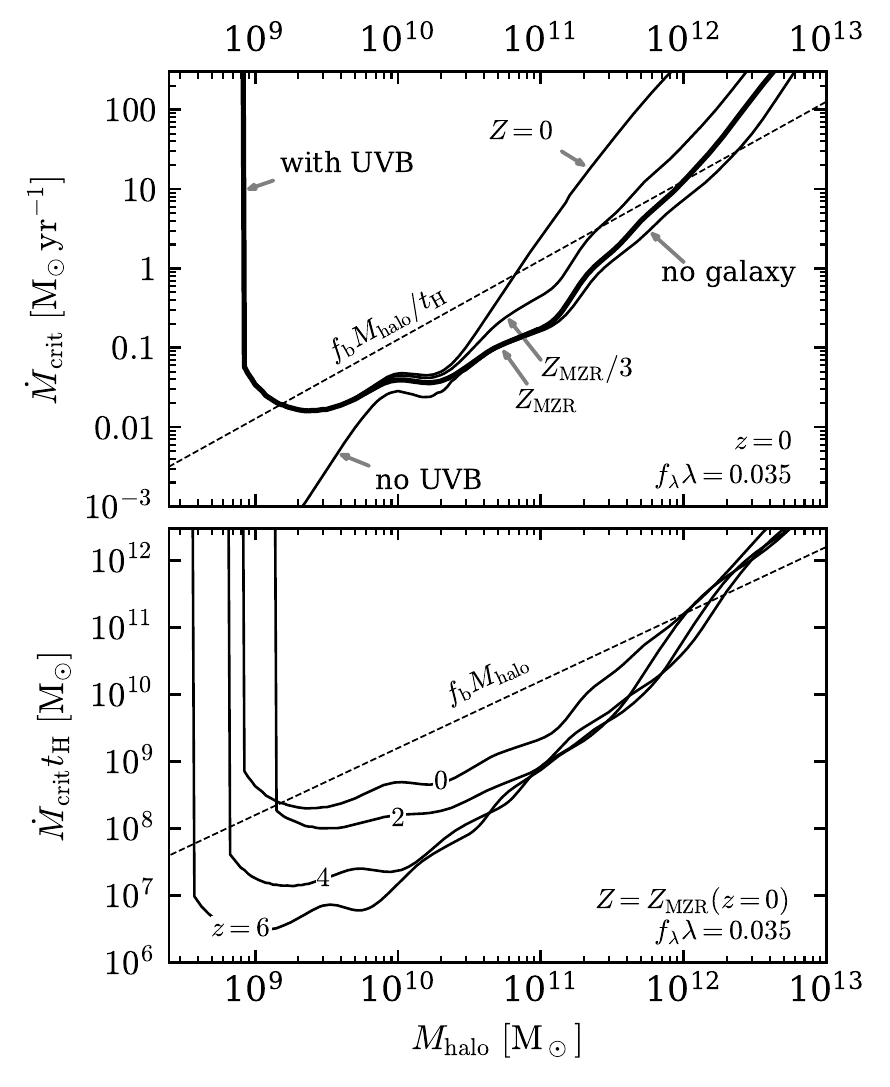}
\caption{The critical accretion rate (eqn.~\ref{e:Mdotcrit num}) versus halo, gas and galaxy  parameters. 
\textbf{(Top)} Halos at $z=0$. 
The thick curve marks $\Mdotcrit$ for a hot gas metallicity (at $\Rcirc\approx0.05\Rvir$) of $\zism$ -- the gas metallicity in the central galaxy based on the observed mass-metallicity relation (eqn.~\ref{e:MZR}). Other curves assume either a different metallicity as noted, or neglect heating by the UVB or the effect of the galaxy on the gravitational potential. 
The dashed line marks the cosmic halo baryon budget divided by the Hubble time. 
\textbf{(Bottom)} The dependence of $\Mdotcrit\thubble$  on redshift, assuming no redshift evolution in the MZR. Heating by the UVB and the gravity of the galaxy are included in the calculation. 
The dashed line marks the cosmic halo baryon budget. 
The intersection of the solid lines with the dashed lines gives the threshold halo mass for the onset of hot mode accretion in baryon-complete halos. 
}
\label{f:Mdotcrit}
\end{figure}

To estimate $\fvc$ we assume an NFW profile for the dark matter and an exponential disk for the galaxy. 
NFW concentration parameters are calculated using the fitting formulas of \cite{Klypin+16}, which are based on the Bolshoi-Planck dark matter simulation.\footnote{\cite{Klypin+16} published concentration parameters of halos up to $z=5.4$; we use the $z=5.4$ values at higher redshifts.} 
For the galaxy mass we use the SMHM from B19, as used above to estimate the gas metallicity. The half mass radius $\Rhalf$ is taken from \cite{Kravtsov13}:
 \begin{equation}\label{e:Rhalf}
 \Rhalf = 0.015 R_{\rm 200c} \approx 0.012\rvir ~,
\end{equation}
where $R_{\rm 200c}$ is the radius enclosing an overdensity of 200 relative to the critical density. 
Since $\fvc$ depends on the mass enclosed within $\Rcirc\approx0.05\rvir$, this size estimate is practically equivalent to assuming the galaxy is a point source, so any $\Rhalf$ up to a factor of $\approx2$ above the estimate in eqn.~(\ref{e:Rhalf}) yields similar results. We then sum the galaxy mass profile with the NFW profile normalized by $1-\Mstar/\Mhalo$, and solve eqn.~(\ref{e:Rcirc}) for $\Rcirc$ and $\fvc$. For $10^{10}-10^{13}\msun$ halos at $z=0$, the value of $\fvc$ is typically larger than unity, and $\Mdotcrit$ increases by a factor of up to two relative to an isothermal calculation with $\fvc=1$. At higher redshifts $2<z<6$ the lower NFW concentration implies that $\fvc<1$, decreasing $\Mdotcrit$ by up to a factor of four relative to an isothermal calculation. 

For completeness, we also calculated a mass profile which accounts for adiabatic contraction of the dark matter due to the galaxy using the {\sc contra} package (\citealt{Gnedin+04}). We find that this effect can only increase $\Mdotcrit$, by a factor of at most two. Observations and cosmological simulations though suggest that this contraction may be negated by dark matter expansion induced by clumpy gas accretion or feedback (\citealt{Dutton+07,Maccio+12,Chan+15}), so we do not consider it further.

In the top panel of Fig.~\ref{f:Mdotcrit} we plot the derived $\Mdotcrit$ for $z=0$ halos, with different assumptions on the calculation of $\Lambda$ and $\fvc$. The thick curve is the fiducial model which assumes a metallicity equal to $\zism$ given by eqn.~(\ref{e:MZR}), heating by the UVB, and includes the effect of the galaxy on $\fvc$. The other curves differ from this fiducial calculation as noted, either by assuming a metallicity equal to a third of the fiducial estimate, no metal contribution to the cooling, no UVB heating, or no effect of the galaxy on the potential. 
An increase in $\Mdotcrit$ with decreasing $Z$ is apparent at $\Mhalo>10^{10.5}\msun$, at which $\Tvir>10^5\K$ and the metals can dominate the cooling.
Heating by the UVB significantly affects $\Mdotcrit$ only at $\Mhalo<10^{10}\msun$ at which the gas temperature is close to the equilibrium temperature. When the gas temperature equals the equilibrium temperature at $\Mhalo\approx10^9\msun$ then $\Mdotcrit$ goes to infinity since $\Lambda$ goes to zero.  Below this threshold there is no net cooling and the cooling flow solutions do not apply.
Note also that \cite{Wiersma+09} did not account for local ionization sources in the galaxy (e.g.~\citealt{Cantalupo10}), which may also decrease $\Lambda$ and increase $\Mdotcrit$. 
Also evident in Fig.~\ref{f:Mdotcrit} is that the galaxy gravity increases $\Mdotcrit$ due to the associated increase in the circular velocity at $\Rcirc$. The largest effect is at $\Mhalo=2.5\cdot10^{12}\msun$ where the SMHM peaks, in which $\Mdotcrit$ increases by a factor of $2.8$. The change in $\Mdotcrit$ due to the galaxy is smaller at lower and higher $\Mhalo$, and almost vanishes at $\Mhalo<10^{11}\msun$ due to the small galaxy mass.

In the bottom panel of Fig.~\ref{f:Mdotcrit} we vary the redshift while keeping $\lambda$ and $Z$ constant. The vertical axis in this panel is $\Mdotcrit\thubble$ where $\thubble$ is the Hubble time. This product gives a characteristic mass associated with accretion at a rate $\Mdotcrit$, and also the gas mass associated with accretion at a rate $\Mdotcrit$ (see below).
Note that with increasing redshift the virial temperature increases for a given halo mass (eqn.~\ref{e:Tvir}), causing the maximum in $\Lambda$ and hence minimum in $\Mdotcrit$ to shift toward lower masses.
At halo masses above the minima in $\Mdotcrit$ the value of $\Mdotcrit\thubble$ depends relatively weakly on redshift. This independence follows since at high $\Mhalo$ metals dominate the cooling and hence $\Mdotcrit\propto \vc^{5.4}\Rcirc$ (eqn.~\ref{e:Mdotcrit num sec2}). Since $\vc\sim\vvir\propto (1+z)^{1/3}$ and $\Rcirc\propto \Rvir\propto (1+z)^{-2/3}$ we get that $\Mdotcrit\thubble\propto (1+z)^{0.13}$, i.e.~$\Mdotcrit\thubble$ is roughly independent of redshift if the metallicity is held constant.
The offset of $\Mdotcrit\thubble$ at high $\Mhalo$ and $z = 0$ relative to at $z\geq2$ is mainly due to the higher concentration of $z = 0$ halos, and hence a higher $\fvc$ as mentioned above.

\begin{figure}
 \includegraphics{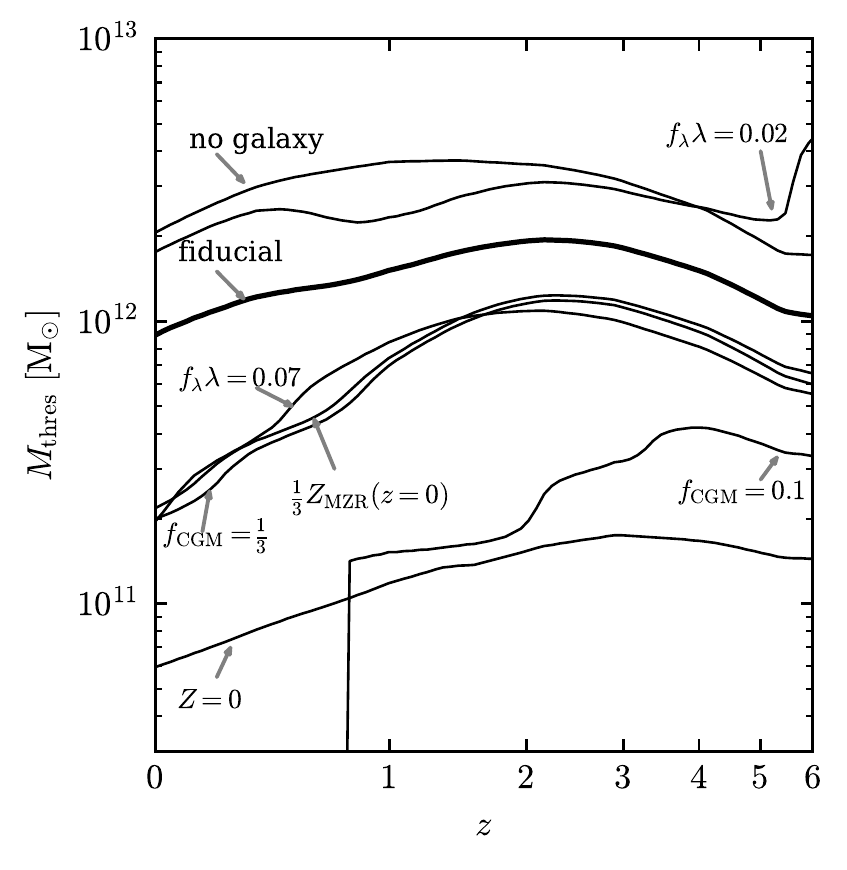}
\caption{The threshold halo mass for hot mode accretion versus redshift, derived by equating the gas mass in the critical solution $\Mdotcrit\thubble$ with a halo gas mass of $f_{\rm CGM}f_{\rm b}\Mhalo$. The thick curve marks $\Mthres$ for a baryon-complete halo ($f_{\rm CGM}=1$), the fiducial gas spin parameter ($f_\lambda\lambda=0.035$), and a hot gas metallicity of $\zism(z=0)$ -- the gas metallicity in the central galaxy based on the observed mass-metallicity relation at $z=0$ (eqn.~\ref{e:MZR}). Other curves assume either a different metallicity, a different gas spin, or different $f_{\rm CGM}$ as noted. The curve marked `no galaxy' neglects the effect of the galaxy on the gravitational potential in the calculation of $\Mdotcrit$. 
}
\label{f:Mthres}
\end{figure}

The product $\Mdotcrit\thubble$ provides a rough estimate of the halo gas mass $\Mgas$ in the critical solution. This follows since in an inflow solution the gas mass equals $\Mdot$ times the crossing time $r/\left|\vr\right|$, which in a cooling flow equals the cooling time at the virial radius (eqn.~\ref{e:self similar v}). In the critical solution $\tcool(\Rvir)\approx\thubble$ (eqn.~\ref{e:WF91 comparison}), so we get that $\Mgas\approx\Mdotcrit\thubble$. Thus, at halo masses where $\Mdotcrit\thubble$ is smaller than the cosmic halo baryon budget $\fb\Mhalo$ (below the dashed lines in Fig.~\ref{f:Mdotcrit}) the critical solution requires the halo to be baryon-depleted, while a baryon-complete halo would have $\Mdot>\Mdotcrit$ and hence be either transonic or entirely supersonic. At halo masses where $\Mdotcrit\thubble>\fb\Mhalo$ (above the dashed lines) we expect $\Mdot<\Mdotcrit$ even in baryon-complete halos and hence the halo gas is expected to be purely subsonic. The intersection of the $\Mdotcrit\thubble$ and $\fb\Mhalo$ curves therefore gives the classic threshold halo mass $\Mthres$ for the onset of hot mode accretion in baryon-complete halos. 

The implied $\Mthres$ derived by equating the gas mass in the critical solution $\Mdotcrit\thubble$ and a gas mass equal to $f_{\rm CGM}\fb\Mhalo$ are plotted in Figure~\ref{f:Mthres}. The thick curve assumes baryon-complete halos ($f_{\rm CGM}=1$), the fiducial gas spin parameter $f_{\lambda}\lambda=0.035$, and a hot gas metallicity given by eqn.~(\ref{e:MZR}), i.e.~equal to the metallicity of the central galaxy assuming no evolution in the mass-metallicity relation with redshift. The threshold halo mass under these assumptions is in the range $\Mthres\approx0.9-2\cdot10^{12}\msun$ at all plotted redshifts. If the hot gas metallicity is a third of this fiducial value $\Mthres$ decreases by a factor of five at $z=0$ and by a smaller factor of two at $z=2$. A similar change in $\Mthres$ is evident if the hot gas mass is a third of the halo cosmic baryon budget or if the angular momentum of the CGM is larger by a factor of two than the fiducial value. Without any metal cooling $\Mthres$ decreases to $0.7-2\cdot10^{11}\msun$. Assuming $f_{\rm CGM}=0.1$ implies that at $z<0.8$ hot mode accretion is possible at at all halo masses. The top curve shows that disregarding the effect of the average galaxy on the potential increases $\Mthres$ by a factor of two. Fig.~\ref{f:Mthres} thus demonstrates that $\Mthres$ can vary significantly according to the gas and galaxy parameters, especially at low redshift.

The derived $\Mthres$ is similar to that found by \cite{DekelBirnboim06} for the same assumed parameters, i.e.~for $z=0$, $Z = Z_\odot$ and a shock radius $\approx 0.1\rvir$ \citeauthor{DekelBirnboim06} derived $\Mthres=2.5\cdot10^{12}\msun$ (see their figure~4), similar to $\Mthres=2\cdot10^{12}\msun$ implied by the `no galaxy' calculation at $z=0$ in Fig.~\ref{f:Mthres}. The weak dependence of $\Mthres$ on redshift in our fiducial parameters is also consistent with their and previous conclusions. Our analysis however emphasizes that physical conditions at $\Rcirc\approx0.05\Rvir$ can significantly change $\Mthres$. The enrichment and depletion of galaxy outskirts by outflows will respectively increase and decrease $\Mthres$. Also, the effect of the galaxy on the gravitational potential decreases $\Mthres$, with a larger decrease for galaxies which are more massive relative to their halo. 
Furthermore, our analysis suggests that $\Mthres$ separates between halos in which the gas is purely subsonic and halos in which the gas is transonic, in contrast with the conclusion of \cite{DekelBirnboim06} that $\Mthres$ separates between purely subsonic and purely supersonic halos (see further discussion below).

\section{Comparison of the critical accretion rate with the star formation rate}\label{s:SFR}

\begin{figure}
 \includegraphics{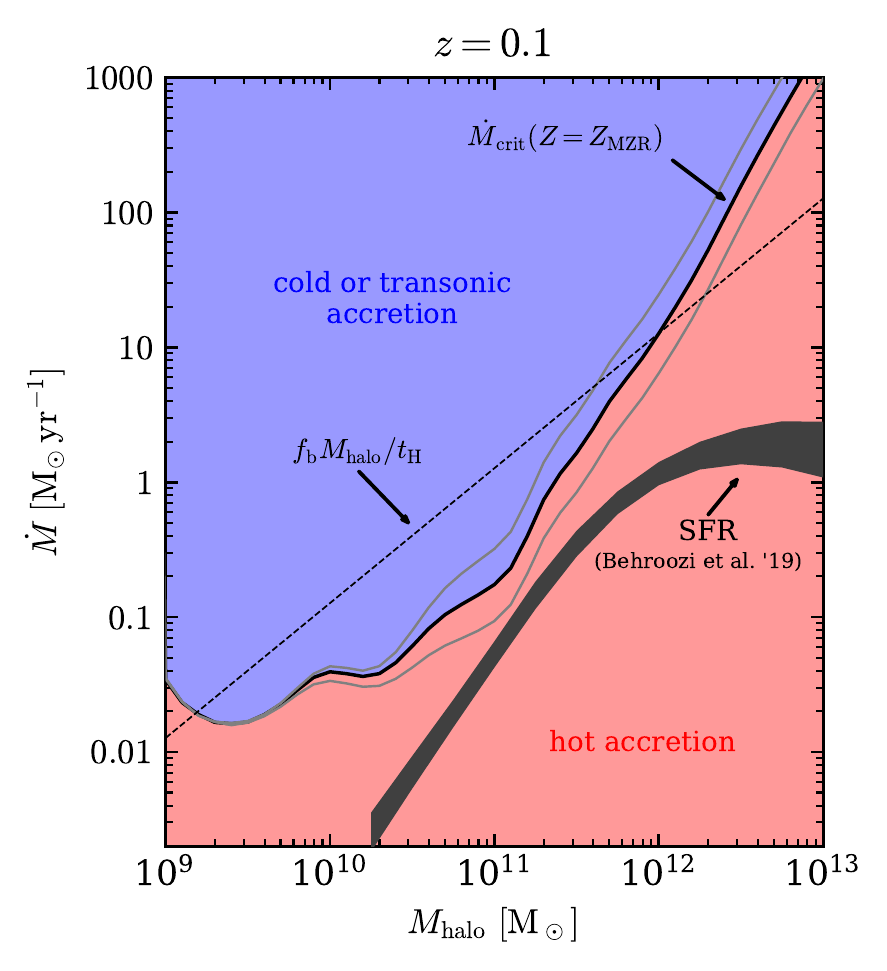}
\caption{Comparison of the critical cooling rate with the mean star formation rate at $z=0.1$.
The thick solid line plots $\Mdotcrit$ (eqn.~\ref{e:Mdotcrit num}), assuming $Z=\zism$ and $f_\lambda\lambda=0.035$. 
Thin grey lines plot $\Mdotcrit$ for a factor of two lower or higher $Z$. For $\Mdot<\Mdotcrit$ we expect the volume-filling phase to be pressure-supported down to the galaxy scale (hot accretion mode, red background), 
while for $\Mdot>\Mdotcrit$ we expect the gas to reach the galaxy with supersonic velocities (cold or transonic accretion modes, blue background). 
The mean SFR for central galaxies derived by B19 is indicated by a gray stripe, where the stripe width marks the uncertainty in their model fit. 
The mean SFR is comparable to or lower than $\Mdotcrit$ at all halo masses. 
}
\label{f:SMHM}
\end{figure}

\begin{figure*}
 \includegraphics{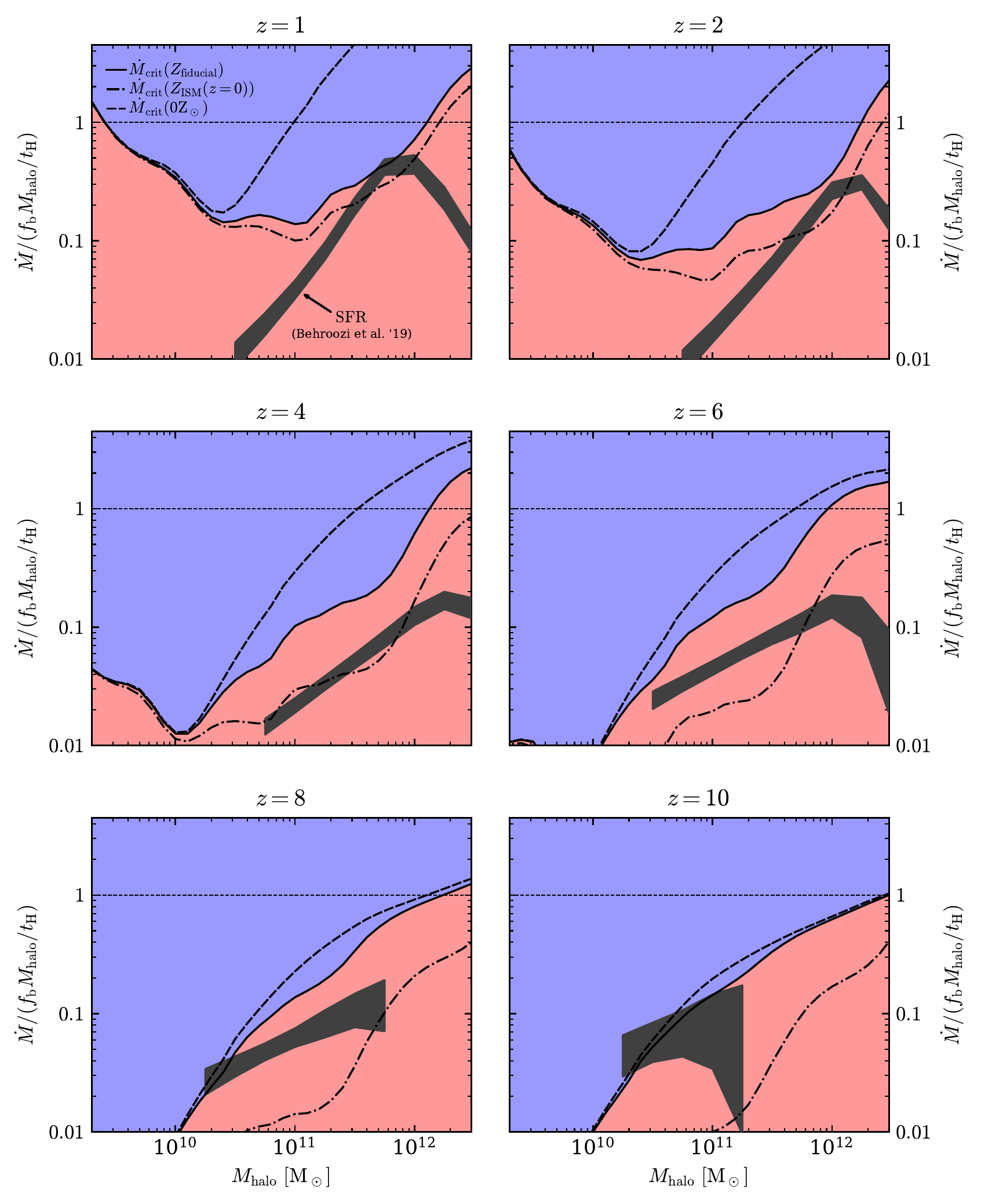}
\caption{Comparison of the critical cooling rate with the mean SFR at different redshifts. To decrease the dynamical range, we normalize the vertical axis by $\fb\Mhalo/\thubble$.
The solid black line plots $\Mdotcrit$ assuming the redshift evolution of the metallicity from Dekel \& Birnboim (2006) given in eqn.~(\ref{e:MZR high z}). Blue and red backgrounds mark respectively hot and cold / transonic accretion modes for the volume-filling phase, for this calculation of $\Mdotcrit$. The two other black lines bracket the uncertainty in $\Mdotcrit$ due to the uncertain metallicity. The thick dashed lines plot $\Mdotcrit$ assuming no metal contribution to the cooling, while the dash-dotted lines assume no redshift evolution in the metallicity. 
Gray bands mark the mean SFR of central galaxies from B19. 
For the fiducial metallicity model $\SFR\lesssim\Mdotcrit$ at all plotted halo masses and redshifts. 
}
\label{f:SMHM all z}
\end{figure*}

In Figure~\ref{f:SMHM} we compare $\Mdotcrit$ in $z=0$ dark matter halos (thick black line) with the average star formation rate (SFR, gray stripe). The value of $\Mdotcrit$ is calculated from eqn.~(\ref{e:Mdotcrit num}) using our fiducial parameters: $f_\lambda\lambda=0.035$, $\fvc$ calculated from an NFW + galaxy profile with $\Mstar(\Mhalo)$ from B19, and the ISM metallicity corresponding to the same $\Mstar$ (eqn.~\ref{e:MZR}). Thin grey lines plot $\Mdotcrit$ assuming  the metallicity is a factor of two lower (top curve) or higher (lower curve) than this fiducial estimate. 
The background colors emphasize the two regimes for how the volume filling phase accretes onto the galaxy, gradual accretion of hot gas if $\Mdot<\Mdotcrit$ and free-fall if $\Mdot>\Mdotcrit$. The average SFR is also taken from B19, and is equal to the time derivative of the SMHM. We plot their mean SFRs for central galaxies (i.e.~excluding satellites), and use the width of the grey stripe to denote the statistical uncertainty in the B19 model fits. The figure demonstrates that the average SFR derived by B19 is less than or comparable to $\Mdotcrit$ at any halo mass. 
As $\Mdotcrit$ is the maximum possible accretion rate of the hot mode, this result suggests that hot mode accretion can in principle dominate the gas supply for star formation in low mass halos.

To extend the comparison of $\Mdotcrit$ with the SFR to high redshift, we assume  $f_\lambda\lambda=0.035$ at all redshifts, motivated by the constant median $\lambda=0.035$ found in the Bolshoi-Planck simulation (\citealt{RodriguezPuebla+16}). The metallicity at high redshift is a major uncertainty. 
As a fiducial model for the metallicity evolution we utilize the scaling suggested by \cite{DekelBirnboim06} based on semi-analytic models:
\begin{equation}\label{e:MZR high z}
 Z(\Mstar,z)= 10^{-s z} Z(\Mstar,z=0)
\end{equation}
with an enrichment rate $s=0.17$. 
This enrichment rate is consistent with the factor of two lower normalization of the mass-metallicity relation at $z\approx 2$ relative to its local value (\citealt{Erb+06, Sanders+15}), and is similar to $s=0.22\pm0.03$ deduced for damped Ly$\alpha$ absorbers (DLAs) at $0<z<5$ by \cite{Rafelski+12}\footnote{The trend of DLA metallicity versus redshift found by \cite{Rafelski+12} does not account for the possible trend of $\Mstar$ with $z$ in their sample, so their quoted value potentially overestimates $s$ as defined in eqn.~(\ref{e:MZR high z}).}. 
The value of $Z(\Mstar,z=0)$ in eqn.~(\ref{e:MZR high z}) is calculated as above using eqn.~(\ref{e:MZR}) for the mass-metallicity relation in the local universe, and using the B19 SMHM to derive $\Mstar$ from $\Mhalo$ and $z$. The same $\Mstar$ is also used for the calculation of $\fvc$, and we assume all the galaxy mass is within $\Rcirc=0.05\rvir$. This latter assumption is consistent with the $\Rhalf\approx0.02\Rvir$ found at $z=0-8$ by \cite{Shibuya+15}.  

Solid lines in Figure~\ref{f:SMHM all z} plot the implied $\Mdotcrit$ using these parameters. Each panel corresponds to a different redshift as noted at the top of the panels. We normalize the vertical axes in this plot by $\fb\Mhalo/\thubble$ in order to decrease the dynamical range, so in this plot the $\Mdotcrit$ curve denotes the required depletion for the onset of hot mode accretion (see section~\ref{s:explore}), while the SFR stripes roughly track the ratio of stellar mass to halo baryon budget (since $\SFR\cdot\thubble/\fb\Mhalo \sim \Mstar/\fb\Mhalo$). 
To bracket the range of $\Mdotcrit$ implied by the uncertainty in metallicity we also plot $\Mdotcrit$ assuming the local mass-metallicity relation holds at higher redshift (i.e., $s=0$), and $\Mdotcrit$ assuming no contribution of metals to the cooling (marked as $Z=0\zsun$, though in practice any $Z\lesssim0.01\zsun$ gives identical results). 
Figure~\ref{f:SMHM all z} demonstrates that for the metallicity evolution rate in eqn.~(\ref{e:MZR high z}), $\SFR\lesssim\Mdotcrit$ at all plotted halo masses and redshifts. The hot mode accretion can thus in principle dominate the gas supply for star formation also in low mass halos at high redshift.

\newcommand{\Mdotbaryons}{\Mdot_{\rm baryons}}
\newcommand{\Mdotavg}{\Mdot_{\rm avg}}
\section{Discussion and Conclusions}\label{s:discussion}

The physical properties of the volume-filling gas phase in dark matter halos are crucial both for understanding the nature of galaxy accretion and for understanding the consequences of feedback (e.g.~\citealt{WhiteRees78, WhiteFrenk91}, BD03, \citealt{DekelBirnboim06, Fielding+17}). 
In this paper we revisit the question of whether this gas phase is predominantly hot and pressure-supported or predominantly cool and free-falling. 
We limit the effect of feedback in our analysis to the possible enrichment and depletion of the halo gas. Absent dynamical effects of feedback (e.g.~heating), hot pressure-supported gas in halos forms a cooling flow. 
We demonstrate that the family of cooling flow solutions separates the physical states of the halo gas into three regimes, according to whether the cooling flow sonic radius $\Rsonic$ is on the scale of the galaxy, on the scale of the halo, or beyond the halo (Fig.~\ref{f:cartoon}). 
The first regime corresponds to the classic hot accretion mode where the flow is subsonic (i.e.~pressure supported) and smooth from the accretion shock down to the galaxy scale. 
The third regime corresponds to the classic cold accretion regime where clumpy gas falls in supersonically from the IGM down to the galaxy without experiencing a strong shock. In the second intermediate regime the gas forms a hot inflow over some range of radii, and then cools out at $\Rsonic$ and free-falls onto the galaxy. This `transonic' regime resembles the classic cold mode in terms of the properties of gas when it accretes onto the galaxy, since the gas reaches the galaxy scale as a cold and free-falling flow (Fig.~\ref{f:flowlines}). However, in terms of coupling with feedback this intermediate scenario may in some aspects more closely resemble the hot mode, due to the existence of a layer of hot and homogeneous pressure-supported gas situated beyond the cold and clumpy free-falling flow. This will presumably depend on where in the halo relative to $\Rsonic$ the feedback energy is deposited. 

In the simulation shown in Figs.~\ref{f:sim meshes}--\ref{f:sim meshes2} the intermediate transonic scenario for the halo gas develops from hydrostatic initial conditions, when the mass inflow rate crosses the critical value of $\Mdotcrit$. 
It is less clear if this scenario can be realized if the gas inflow is supersonic at large radii as is often the case in the cosmological context. 
Indeed, a sonic transition at intermediate radii in the halo is not seen in the idealized simulations of \cite{BirnboimDekel03} where the halo gas is initially inflowing supersonically. We argued in section~\ref{s:BD03} that this difference could be due to the lack of a source of strong shocks in the outer halo in the \citeauthor{BirnboimDekel03} simulations. It would thus be interesting to check whether this intermediate regime materializes in setups which include outflows that shock agains supersonic inflows at large scales (e.g.~\citealt{Fielding+17}) and in the more realistic conditions in cosmological simulations. The latter can include strong feedback at high redshift that `pre-heats' the gas, stifling later supersonic inflows and more closely resembling the hydrostatic initial conditions used in this work (e.g.~Figs.~\ref{f:sim meshes}--\ref{f:sim meshes2}).  We leave addressing this question and deriving the implications of this possible new accretion regime of halo gas to future work. 

We demonstrate that hot mode accretion is possible only if $\Rsonic<\Rcirc\approx0.05\rvir$, because this condition determines when the gas is virialized and roughly hydrostatic down to galaxy scales. This condition on $\Rsonic$ is equivalent to the condition $\tcool\gtrsim\tff$ at $r=\Rcirc$ (eqn.~\ref{e:tcool to tff}), and can be cast as a maximum accretion rate in the hot mode $\Mdotcrit$ (eqns.~\ref{e:Mdotcrit}, \ref{e:Mdotcrit num sec2}). 
We emphasize that `hot' corresponds to the virial temperature, which is relatively low for low-mass halos.  
We explore the dependence of $\Mdotcrit$ on halo mass, redshift, and gas metallicity in Fig.~\ref{f:Mdotcrit}. We find that in halos where metals dominate the cooling the product $\Mdotcrit\thubble$ is roughly independent of redshift if the metallicity is held constant (Fig.~\ref{f:Mdotcrit}).

The classic threshold halo mass for the onset of hot mode accretion $\Mthres$ can be derived by noting that the halo gas mass for accretion at a rate $\Mdotcrit$ is $\approx\Mdotcrit\thubble$ (section~\ref{s:explore}). Since in cooling flows the accretion rate increases with gas mass and density (Fig.~\ref{f:steady-state no AM}), hot mode accretion is expected when the halo gas mass is $\lesssim\Mdotcrit\thubble$. For baryon-complete halos $\Mthres$ can thus be derived from the condition $\Mdotcrit=\fb\Mhalo/\thubble$. Assuming halos are indeed baryon complete, we find $\Mthres \sim 10^{12}\msun$ roughly independent of redshift if the metallicity is held constant (Fig.~\ref{f:Mthres}). This result is comparable to the calculations of BD03 for the formation of a stable accretion shock near $\Rcirc$ for the same parameters. As our derivation does not assume the gas was heated to $\sim\Tvir$ in a single shock, our results suggest that the condition for hot mode accretion derived by BD03 apply more generally. 
We also show that when accounting for the
gravitational effects of the average galaxy, $\Mthres$ decreases by a factor of $\approx 2$ (Fig.~\ref{f:Mthres}).  This demonstrates that the existence of hot mode accretion depends not only on the properties of the halo but also on the properties of the galaxy.
Moreover, we showed 
that if the halo gas mass is depleted relative to its baryon budget 
such that the cooling and accretion rates are smaller than $\Mdotcrit$, then hot mode accretion would be relevant also in halos with $\Mhalo<\Mthres$.

The conclusion that hot mode accretion is determined by conditions at the galaxy scale implies that the 
relevant metallicity for calculating $\Mdotcrit$ is the metallicity of the hot gas just outside the galaxy, which is potentially higher than at larger scales due to more intense enrichment by outflows. 
Also, while our calculations neglect possible deviations from spherical symmetry induced by cosmological filaments, we expect these to affect our results regarding the nature of accretion from the volume-filling phase only if filaments retain their identity down to the galaxy scale. Cosmological simulations currently differ on whether this is indeed the case, or whether instead the filaments dissolve farther out in the halo (e.g., \citealt{Keres+05,Ceverino+10,FaucherGiguere+11, Nelson+13, Danovich+15}, see also \citealt{Mandelker+16,Mandelker+19,Padnos+18}).

Our analysis assumes steady-state conditions, while various physical processes associated with galaxy formation, such as the growth of the background potential, bursty stellar feedback (e.g.~\citealt{Muratov+15}), and clumpy accretion,  could drive the system away from steady state. An interesting question is thus what are the relevant timescales on which steady state can be achieved? Our results suggest that the relevant timescale for determining the nature of accretion is the dynamical time of the galaxy. 
The importance of this timescale emerges from the critical solution, in which $\tcool\approx\tff$ at $r\approx\Rcirc$. Since in any cooling flow solution $\tcool\approx\tflow\equiv r/|v_r|$ (eqn.~\ref{e:self similar v}), the critical solution satisfies
\begin{equation}\label{e:timescales}
 \tcool(\Rcirc) \approx \tflow(\Rcirc) \approx \tff(\Rcirc) = \frac{2f_\lambda\lambda\rvir}{\vc} ~,
\end{equation}
where the last equality follows from eqns.~(\ref{e:tff}) and (\ref{e:Rcirc}). Note that this relation differs from the predictions of several feedback-regulation models in which $\tcool$ in the halo is regulated to some factor of $\tff$ (e.g., \citealt{Sharma+12, Voit+17}), since in these models $\tflow\gg\tcool$ due to heating by feedback, in contrast with $\tflow\approx\tcool$ in the cooling flow solution. Equation~(\ref{e:timescales}) implies that the relevant timescales for the onset of hot mode accretion are a factor of $(\sqrt{2}f_\lambda\lambda)^{-1}\sim20$ shorter than the halo dynamical time, or a factor of $\sim100$ shorter than the Hubble time at the corresponding redshift. We thus expect our results to be roughly valid as long as other processes change the relevant physical conditions on timescales longer than this characteristic value. Moreover, the fact that this timescale is relatively short implies that the nature of accretion can be  determined by transient processes, if the transient conditions last longer than the galaxy dynamical time. For example, if a burst of feedback depletes gas in the galaxy vicinity such that  $\Mdot$ drops below $\Mdotcrit$, then the remaining gas may accrete in the hot mode even if the accretion rate averaged over longer timescales is larger than $\Mdotcrit$.

Figures~\ref{f:SMHM} and \ref{f:SMHM all z} plot the average SFR in dark matter halos empirically derived by \cite{Behroozi+18} based on predictions from dark matter-only simulations and observational constraints. These figures show that the average SFR is lower than $\Mdotcrit$ at almost all halo masses and redshifts, for the fiducial metallicity evolution discussed in section~\ref{s:SFR}. 
It is unclear if this result is a coincidence, or indicates a physical connection between $\Mdotcrit$ and the SFR in low mass halos. However, we have shown that hot mode accretion and $\Mdotcrit$ may be relevant also to low mass halos if they are sufficiently depleted of baryons. It would thus be valuable to explore scenarios in which $\Mdotcrit$ provides a physical upper limit to the SFR in all halos at all times. 
How could this be the case? 
Due to the different nature of accretion and different consequences of feedback according to the state of the halo gas, it is plausible that the star formation efficiency $\SFR/\Mdot$ during the hot accretion phase differs significantly from the SF efficiency during the phase where gas reaches the galaxy in free-fall. In a low-mass halo where gas accretes onto the galaxy via the hot mode for some fraction of the time, and the SF efficiency during this hot mode phase is high while it is low in the free-fall phase due to strong winds, the SFR would tend to be $\lesssim\Mdotcrit$, since during the hot phase $\Mdot\lesssim\Mdotcrit$. Simulations of low mass halos which include star formation and feedback could test if such a scenario is realized.

Last, since $\Mdotcrit$ is determined by physical properties at the galaxy scale it can be estimated from observations of galaxy properties, and then compared to the SFR (or other properties) of individual galaxies. This is in contrast with the statistical modelling required to derive average SFR and $\Mdotcrit$ in dark matter halos using techniques such as abundance matching (as in section~\ref{s:SFR} above). It would be interesting to derive the relation between SFR and $\Mdotcrit$ on a galaxy-by-galaxy basis and for different galaxy subtypes. 
This may provide new insights into the importance of the hot accretion mode for fuelling and/or quenching star formation, as well as the origin of galaxy scaling relations involving
parameters determining $\Mdotcrit$, such as the \cite{TullyFisher77}~relation between $v_{\rm c}$ and stellar mass.

\section*{Acknowledgements}
JS is supported by the CIERA Postdoctoral Fellowship Program. DF is supported by the Flatiron Institute, which is supported by the Simons Foundation. CAFG is supported by NSF through grants AST-1517491, AST-1715216, and CAREER award AST-1652522, by NASA through grants NNX15AB22G and 17-ATP17-0067, by STScI through grants HST-GO-14681.011, HST-GO-14268.022-A, and HST-AR-14293.001-A, and by a Cottrell Scholar Award from the Research Corporation for Science Advancement. EQ was supported in part by a Simons Investigator Award from the Simons Foundation and by NSF grant AST-1715070.


\appendix

\section{The Bernoulli parameter in the presence of radiative losses}\label{a:Bernoulli}

When accounting for radiative losses, energy conservation can be stated as
\begin{equation}
 \frac{\der \epsilon}{\der t} = -\frac{P\der \rho^{-1}}{\der t} - q = \frac{P}{\rho}\frac{\der \ln \rho}{\der t} - q ~,
\end{equation}
where $\epsilon$ and $q$ are the specific thermal energy and specific luminosity, and the other variables have their usual meaning. 
For a spherical steady-state flow $\der/\der t = v_r\der/\der r$ so we get
\begin{equation}
 v_r\left(\frac{\der \epsilon}{\der r} -\frac{P}{\rho}\frac{\der \ln \rho}{\der r}\right) = -q ~.
\end{equation}
Using $\der \ln\rho = \der\ln P-\der\ln\epsilon$ and $P/\rho = (\gamma-1)\epsilon$ then gives
\begin{equation}\label{e:app}
 v_r\left(\gamma\frac{\der \epsilon}{\der r} -\frac{1}{\rho}\frac{\der P}{\der r}\right) = -q ~.
\end{equation}
Using the momentum equation~(\ref{e:momentum1}) we then arrive at eqn.~(\ref{e:acc vs. cool}):
\begin{equation}
 v_r\frac{\der}{\der r}\left(\frac{1}{2}v_r^2+\gamma\epsilon+\Phi\right) = -q ~,
\end{equation}

\section{Formation of cooling flows from hydrostatic initial conditions}\label{a:collapse}

\begin{figure*}
 \includegraphics{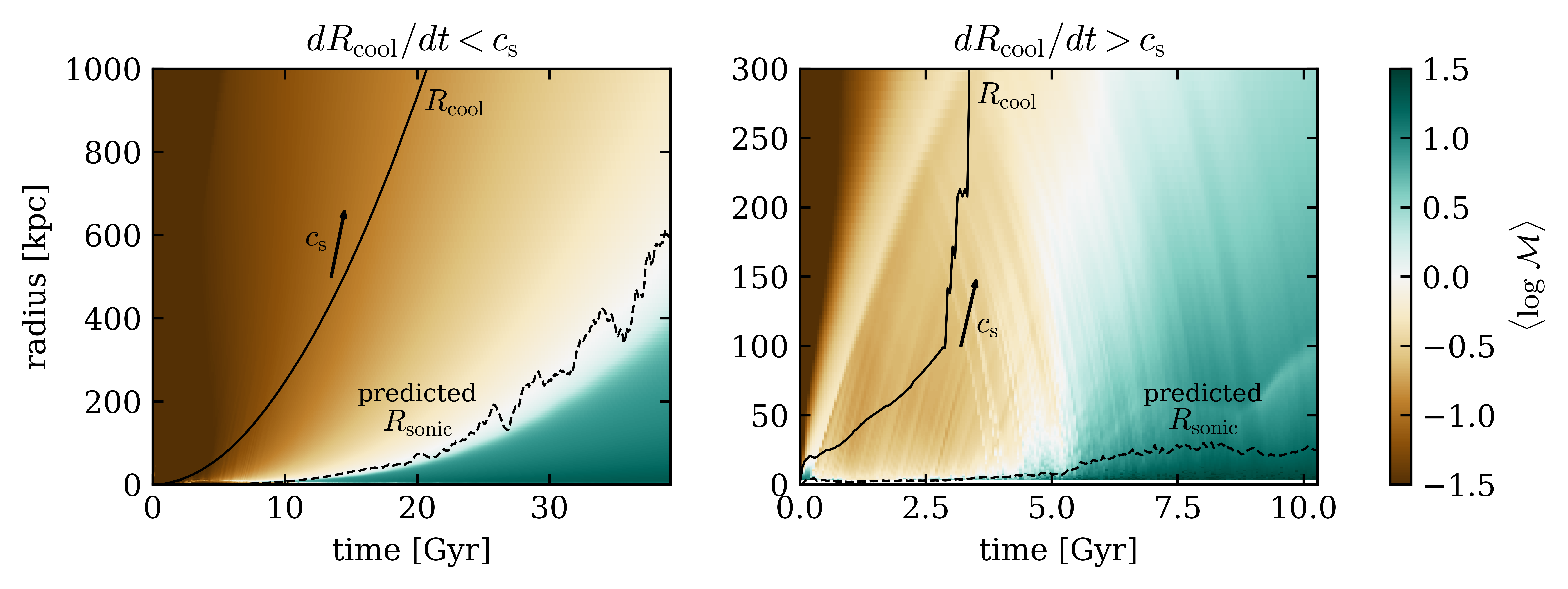}
\caption{
Formation of a cooling flow versus collapse into a supersonic flow. The \textbf{left panel} plots the simulation used in this work, while the \textbf{right panel} plots the high density $10^{12}\msun$ simulation from Paper I. Initial conditions are hydrostatic in both simulations, while background color maps plot the shell-averaged Mach number at each radius and time. The solid black lines plot the cooling radii. The dashed lines plot the predicted sonic radii in different snapshots, based on $\Mdot$ in each snapshot and the relation between $\Mdot$ and $\Rsonic$ in cooling flows (eqn.~\ref{e:Rsonic}). In the left panel $\Rcool$ expands slower than the local sound speed (indicated by the slope of the $c_{\rm s}$ arrow), and a transonic flow forms with the predicted $\Rsonic$ roughly equal to the actual $\Rsonic$ in the simulation (white contour). 
In the right panel $\Rcool$ expands faster than the sound speed after $t=3\Gyr$, and the halo gas collapses into a purely supersonic flow with $\Rsonic\rightarrow\infty$.
}
\label{f:collapse}
\end{figure*}

Figure~\ref{f:collapse} plots the shell-averaged Mach number as a function of radius and time in the simulation used in this work (left panel) and in the high density $10^{12}\msun$ simulation from Paper I (right panel, see also figures 5 and 9 in Paper I). Also plotted are $\Rcool$ and $\Rsonic(\Mdot)$ predicted by the cooling flow solutions (eqn.~\ref{e:Rsonic}), based on $\Mdot(t,r=20\kpc)$ measured in each snapshot. In the simulation used in this work, $\Rcool$ expands slower than the local sound speed, and the flow converges onto the steady-state solutions with the predicted $\Rsonic$ matching the actual $\Rsonic$ in the simulation, as discussed in section~\ref{s:hydro}. In contrast in the Paper I simulation $\Rcool$ expands faster than the sound speed at $t\gtrsim3\Gyr$, and within $\approx2\Gyr$ the halo gas collapses into a purely supersonic flow with $\Rsonic\rightarrow\infty$. Figure~\ref{f:collapse} thus suggests that $d\Rcool/d t <\cs$ is a necessary condition for the convergence of initially hydrostatic gas onto the family of cooling flow solutions discussed in this work. The same condition was imposed by B89 in order to derive their self-similar cooling-wave solutions. 
It is possible however that in realistic systems a purely supersonic inflow will shock against outflows from the galaxy and form a cooling flow (see section~\ref{s:BD03}). 

\label{lastpage}

\end{document}